\documentclass[twocolumn]{aastex63}
\usepackage{lineno}

\graphicspath{{./}{figures/}}

\received{}
\revised{}
\accepted{}

\submitjournal{ApJ}

\shorttitle{SpaceTiger}
\shortauthors{Maksyutenko et al.}

\begin{document}

\title{Formation of $c$-C$_6$H$_5$CN ice using the SPACE TIGER experimental setup}

\correspondingauthor{Rafael Mart\'in-Dom\'enech}
\email{rafael.martin$\_$domenech@cfa.harvard.edu}

\author{Pavlo Maksyutenko}
\altaffiliation{Woefully, Pavlo Maksyutenko passed away while this project was in progress. It has been completed by Rafael Mart\'in-Dom\'enech and collaborators on his behalf.}
\affiliation{Center for Astrophysics $|$ Harvard \& Smithsonian\\
60 Garden St., Cambridge, MA 02138, USA}

\author[0000-0001-6496-9791]{Rafael Mart\'in-Dom\'enech}
\affiliation{Center for Astrophysics $|$ Harvard \& Smithsonian\\
60 Garden St., Cambridge, MA 02138, USA}

\author[0000-0001-6947-7411]{Elettra L. Piacentino}
\affil{Center for Astrophysics $|$ Harvard \& Smithsonian\\
60 Garden St., Cambridge, MA 02138, USA}

\author[0000-0001-8798-1347]{Karin I. \"Oberg}
\affil{Center for Astrophysics $|$ Harvard \& Smithsonian\\
60 Garden St., Cambridge, MA 02138, USA}

\author[0000-0003-2761-4312]{Mahesh Rajappan}
\affil{Center for Astrophysics $|$ Harvard \& Smithsonian\\
60 Garden St., Cambridge, MA 02138, USA}


\begin{abstract}
Benzonitrile ($c$-C$_6$H$_5$CN) has been recently detected in cold and dense regions of the interstellar medium (ISM), where it has been used as a signpost of a rich aromatic organic chemistry that might lead to the production of polycyclic aromatic hydrocarbons (PAHs).  
One possible origin of this benzonitrile is interstellar ice chemistry involving benzene ($c$-C$_6$H$_6$) and nitrile molecules (organic molecules containing the $-$C$\equiv$N group). 
We have addressed the plausibility of this $c$-C$_6$H$_5$CN formation pathway through laboratory experiments 
using our new setup SPACE TIGER. 
The SPACE TIGER experimental setup is designed to explore the physics and chemistry of interstellar ice mantles using laser-based ice processing and product detection methods.  
%
%
We have found that $c$-C$_6$H$_5$CN is formed upon irradiation of $c$-C$_6$H$_6$:CH$_3$CN binary ice mixtures with 2 keV electrons and Lyman-$\alpha$ photons at low temperatures (4$-$10 K).
Formation of $c$-C$_6$H$_5$CN was also observed when $c$-C$_6$H$_6$ and CH$_3$CN were embedded in a CO ice matrix, but it was efficiently quenched in a H$_2$O ice matrix. 
The results presented in this work imply that interstellar ice chemistry involving benzene and nitrile molecules could contribute to the formation of the observed benzonitrile only if these species are present on top of the ice mantles or embedded in the CO-rich ice layer, instead of being mixed into the H$_2$O-rich ice layer.  

\end{abstract}


\section{Introduction} \label{sec:intro}


Around 10\%$-$25\% of the carbon present in the interstellar medium (ISM) is estimated to be in the form of polycylic aromatic hydrocarbons \citep[PAHs,][]{dwek97,chiar13}.   
PAHs are the proposed carriers of the unidentified infrared (UIR) emission bands that are ubiquitously detected in the ISM \citep{low84,leger84,allamandola85}. 
As a significant reservoir of carbon, these organic, aromatic macromolecules could play an important role in the interstellar formation of other complex organic species with prebiotic interest. 
However, it is currently unclear how the PAHs form in the ISM \citep[][]{joblin18}.   
In order to account for the presence of these molecules in both the diffuse and the dense regions of the ISM, two formation scenarios have been proposed in the literature. 
PAHs could form upon destruction of larger carbonaceous solids 
in the hot, dense outflows of carbon-rich evolved stars \citep[e.g.,][]{pilleri15,martinez20}, as well as in the diffuse ISM \citep{berne15}.
In the second scenario, 
PAHs are formed from smaller hydrocarbons \citep{woods02,cernicharo04} and simpler aromatic molecules. 
This bottom-up scenario would be the only available formation mechanism in cold and dense regions of the ISM where carbonaceous particles are shielded from destruction by the interstellar UV field and/or shock waves \citep{mcguire18}. 

In the bottom-up formation scenario, the cyclization of the small hydrocarbons leading to the first aromatic hydrocarbon \citep[either benzene, $c$-C$_6$H$_6$, or the phenyl radical, $c$-C$_6$H$_5^.$,][]{ilsa20} is expected to be the limiting step for the formation of PAHs \citep{cherchneff92,tielens97,kaiser15}. 
Therefore, understanding the formation of $c$-C$_6$H$_6$ and characterizing its presence in the dense regions of the ISM is highly relevant for the study of the PAHs formation process. 
%
%
However, the detection of this molecule in the interior of cold and dense clouds by means of radio astronomy has been challenging due to the lack of a permanent dipole moment. 
Instead, radio astronomers have been able to detect the N-bearing aromatic heterocycle 
benzonitrile ($c$-C$_6$H$_5$CN, dipole moment = 4.5 D) 
toward the TMC-1 dense core of the Taurus Molecular Cloud \citep{mcguire18}, and more recently toward four other prestellar cores  \citep{burkhardt21a}. 
Benzonitrile is the only benzene derivative firmly detected in the ISM thus far \citep[a tentative detection of phenol toward the high mass star-forming region Orion KL was reported in][]{kolesnikova13}. 
The recent detection of benzonitrile has been followed by the subsequent detection of the single-ring aromatic species
cyclopentadiene \citep[$c$-C$_5$H$_6$,][]{cernicharo21}, along with  
1- and 2- cyanocyclopentadiene \citep[C$_5$H$_5$CN,][]{mccarthy20,lee21};  
as well as the two-ring aromatic molecules 
indene \citep[$c$-C$_9$H$_8$,][]{burkhardt21},  
and 1- and 2-cyanonaphtalene \citep[C$_{10}$H$_7$CN,][]{mcguire21},  
also toward TMC-1. 

In the light of these detections, previous works have suggested that $c$-C$_6$H$_5$CN could be used as a proxy to characterize the presence of $c$-C$_6$H$_6$ in the interior of dense clouds, since it is proposed to form readily in the gas phase through a barrierless neutral-neutral reaction between $c$-C$_6$H$_6$ and CN \citep{woods02,trevitt10}. 
%
The rate coefficient for the $c$-C$_6$H$_6$ + CN gas phase reaction has been recently found to be 
constant
in the 15 K $-$ 295 K range, consistent with a barrierless reaction \citep{ilsa20}. 
However, this measured rate coefficient leads to an underprediction of the gas-phase $c$-C$_6$H$_5$CN by astrochemical models, compared to that observed toward TMC-1 \citep{mcguire18}. 
%
While \citet{ilsa20} suggested that this discrepancy was probably due to an underestimation of the initial $c$-C$_6$H$_6$ abundance in the model, the difference between the predicted $c$-C$_6$H$_5$CN gas-phase abundance and the observed value could also be explained by missing $c$-C$_6$H$_5$CN formation pathways in the theoretical models \citep{mcguire18}, including formation in interstellar ice mantles. 

Interstellar ice mantles are grown on the surface of dust grains in the coldest and densest regions of the ISM \citep{boogert15}, such as the TMC-1 dense core, where the bottom-up scenario is expected to dominate the formation of PAHs. 
Under such temperature and density conditions, a fraction of the interstellar $c$-C$_6$H$_6$ could be present in the solid phase \citep[either by accretion from the gas-phase, or by formation \textit{in situ}, see, e.g.,][]{abplanalp19}. 
The subsequent evolution of these ice mantles is characterized by a rich chemistry induced by their energetic processing,   
%
which is driven by the secondary keV electrons and the secondary UV field arising from the interaction of the cosmic rays with the ice molecules \citep{bennett05}, or the gas-phase H$_2$ molecules \citep{cecchi92,shen04}, respectively. 
Recent experimental simulations have revealed that photoprocessing of $c$-C$_6$H$_6$:HCN ice mixtures with $\lambda$ $>$ 200 nm UV photons leads to the formation of several nitriles, including $c$-C$_6$H$_5$CN \citep{mouzay21}. 
%
%
Here, we have expanded on this work and tested the solid-state formation of $c$-C$_6$H$_5$CN upon energetic processing with both Ly-$\alpha$ ($\lambda$ = 121.2 nm) VUV photons and 2 keV electrons of ice samples containing $c$-C$_6$H$_6$ and acetonitrile (CH$_3$CN), at low temperatures relevant for dense cloud conditions ($\sim$4$-$10 K).  

We have also used these laboratory simulations of the $c$-C$_6$H$_5$CN formation in interstellar ices as a case of study to present the new SPACE TIGER\footnote{Surface Photoprocessing Apparatus Creating Experiments To Investigate Grain Energetic Reactions} experimental setup, designed to explore the physics and chemistry of ice samples analog to the ice mantles detected in the ISM. 
%
%
\S \ref{sec:exp} presents the SPACE TIGER setup along with its main features and capabilities. The laboratory results on the $c$-C$_6$H$_5$CN ice formation are presented in \ref{sec:results} and discussed in \ref{sec:disc}. 
Finally, the main conclusions are summarized in \ref{sec:conc}. 

\begin{figure*}
    \centering
    \includegraphics[width=18cm]{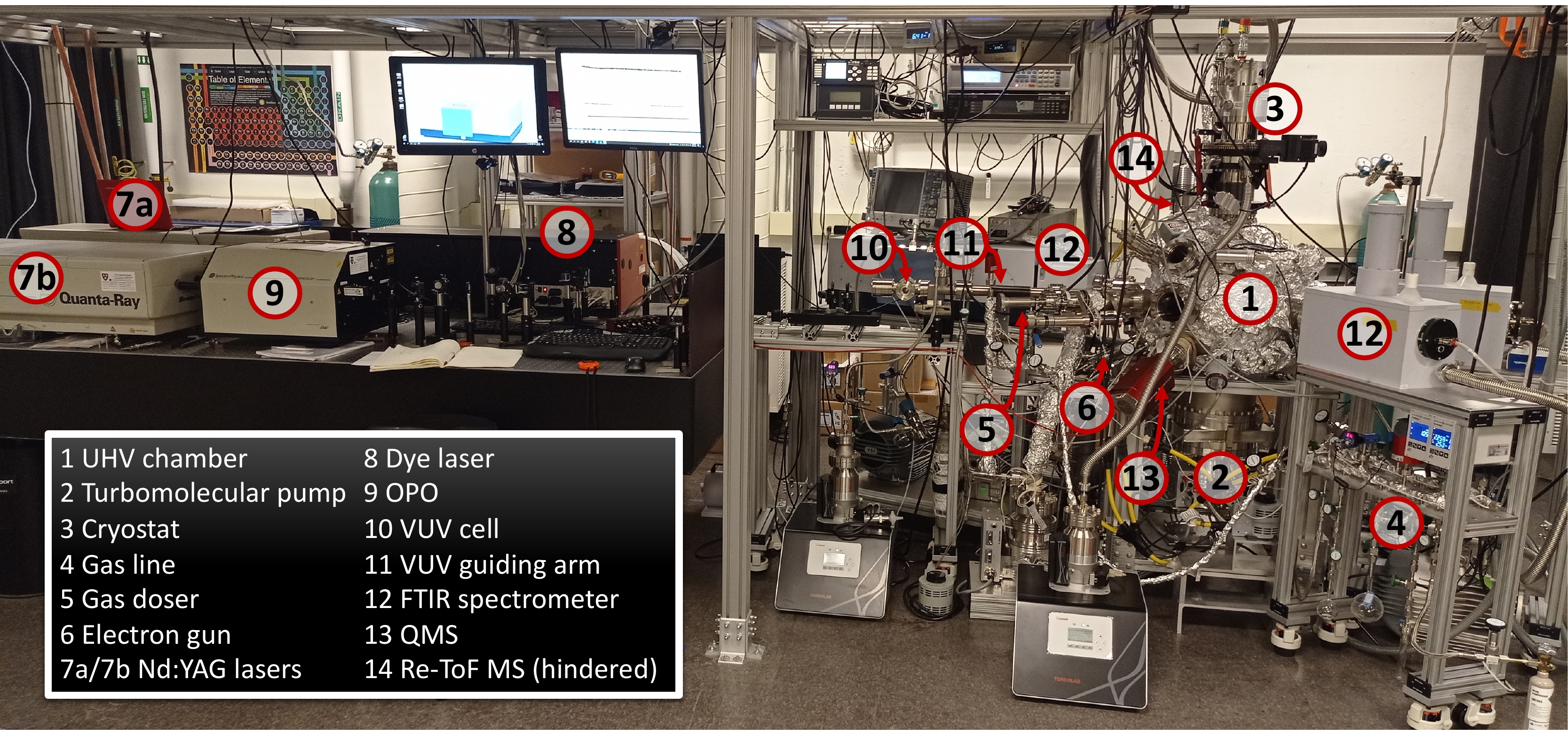}
    \caption{Picture of the SPACE TIGER experimental setup with the most important components highlighted. The gas line and doser assemblies are shown in detail in Fig. \ref{fig:ST_gasline}.}
    \label{fig:ST_picture}
\end{figure*}

\section{Experimental setup}\label{sec:exp}

The experiments were performed in the new SPACE TIGER setup (Fig. \ref{fig:ST_picture}). 
This setup consists of a custom made 500 mm diameter spherical stainless steel, ultra-high-vacuum (UHV) chamber (Pfeiffer Vacuum).  
Different instruments and components 
attached to the chamber through its multiple ports (described below) 
allow for the simulation and analysis of the energetic processing of interstellar ice analogs. 
The UHV chamber achieves a base pressure of $\sim$2 $\times$ 10$^{-10}$ Torr at room temperature by means of a water cooled, magnetically levitated Leybold MAG W2200 iP turbomolecular pump (with a pumping capacity of 2100 l s$^{-1}$) connected to the bottom flange of the UHV chamber, and a Leybold MAG W600 iP pump (with a pumping capacity of 600 l s$^{-1}$), connected to the flight tube of the reflectron time-of-flight mass spectrometer (Re-ToF MS) described in Sect. \ref{sec:tof}. 
Both pumps are backed by another magnetically levitated Leybold MAG W300 iP turbomolecular pump (300 l s$^{-1}$), 
which is backed by an Edwards scroll pump nXDs 15i (4 l s$^{-1}$). 
The pressure in the UHV chamber is measured by an IONIVAC IM 540 extractor pressure gauge.

\subsection{Ice sample holder and preparation}\label{sec:dosing}

\begin{figure}
    \centering
    \includegraphics[width=2.cm]{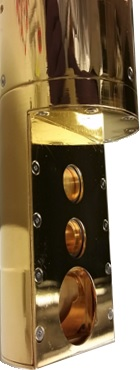}
    \caption{Picture of the three-port OFHC copper sample holder surrounded by a gold-plated radiation shield. The three opennings of the radiation shield correspond to (from top to bottom) the phosphorus screen, the CsI substrate, and the copper substrate.}
    \label{fig:ST_sampleholder}
\end{figure}

Ice samples are grown in situ either on a 12 mm diameter copper substrate or on a 12.7 mm diameter CsI substrate. 
Only the internal 10 mm diameter surface of the CsI substrate is exposed by the gold-plated radiation shield described below. 
The copper substrate is used when the ice samples are irradiated with electrons (Sect. \ref{sec:electron}) and/or when the Re-ToF MS (Sect. \ref{sec:tof}) is used to detect the desorbing ice molecules, since the sample needs to be grounded in those cases.
The CsI substrate cannot be used in any experiments involving electron irradiation or Re-ToF mass spectrometry due to substrate charging. 
The two substrates are mounted on a three-port, oxygen-free high thermal conductivity (OFHC) copper sample holder 
located at the center of the UHV chamber. 
A picture of the sample holder is shown in Fig. \ref{fig:ST_sampleholder}.
The sample holder also incorporates a 10 mm diameter 
phosphorous screen, used to visualize the spot size of the 2 keV electron and VUV photon beams (see Sections \ref{sec:electron} and \ref{sec:uv}). 
The sample holder is surrounded by a gold-plated radiation shield, and connected to the first stage of a closed-cycle He cryostat (Model DE210B-g, Advance Research Systems, Inc.) powered by an ARS-10kW compressor. The radiation shield has openings only in front of the three sample positions. 
With this arrangement, the sample holder can be cooled down to 4.3 K, as verified by a calibrated Si diode sensor 
located close to the copper substrate, 
and monitored by a temperature controller (Lakeshore Model 336).

\begin{figure}
    \centering
    \includegraphics[width=8cm]{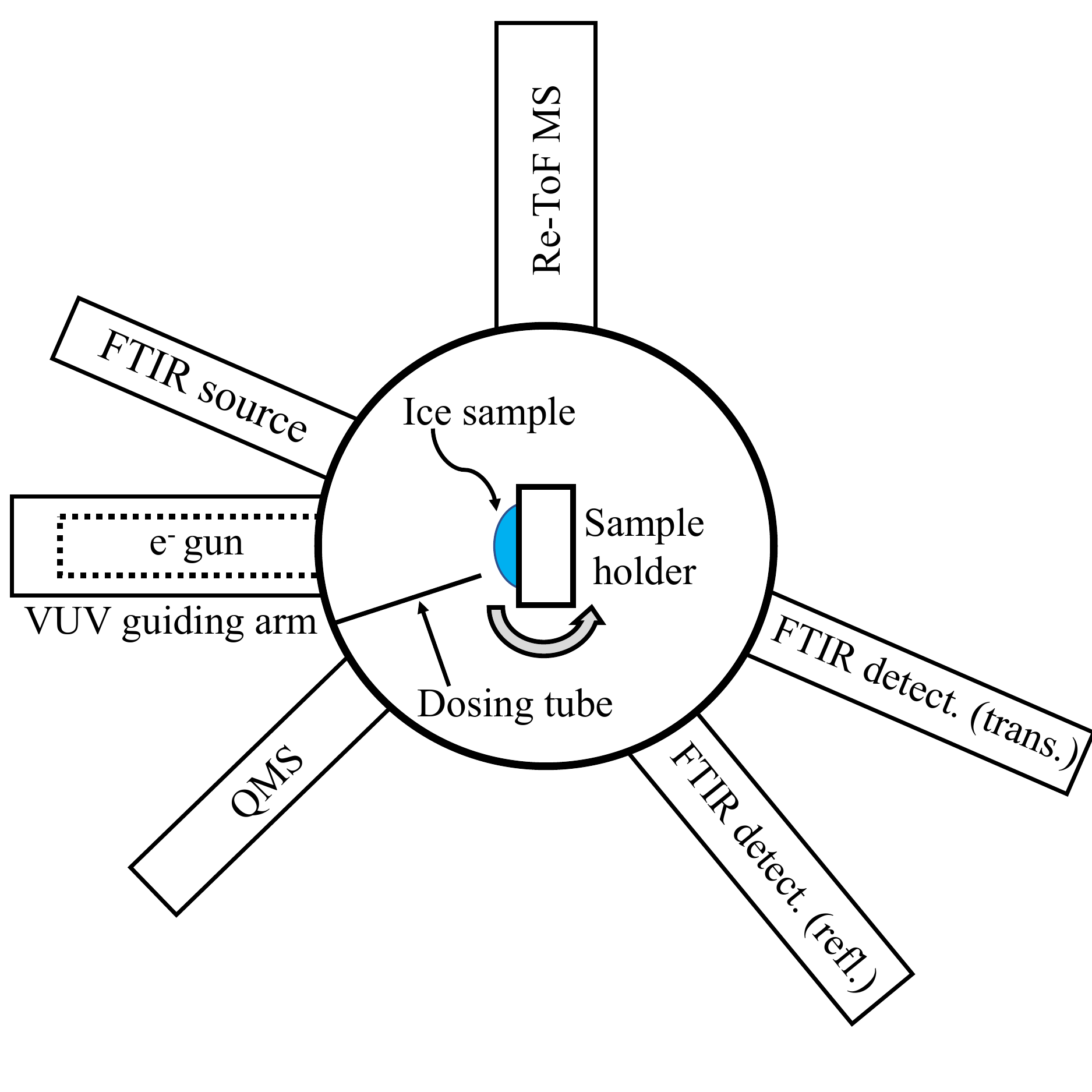}
    \caption{Schematic representation of the top view (xy plane) of the SPACE TIGER UHV chamber, including the rotatory sample holder at the center of the chamber and the different attached instruments.}
    \label{fig:ST_sketch}
\end{figure}

The cryostat with the connected sample holder is mounted on the top flange of the UHV chamber with a differentially pumped (using a Leybold Turbolab 90i pump station with a 90 ls$^{-1}$ pumping capacity) UHV rotatory seal (LMM \& RNN-400/TM/FA) that allows 360$^\circ$ rotation in the xy plane. 
In addition, a 10 cm z-axis translation stage (Thermionics ZC-B600C-T600TM-4.00-3) is used to position the substrate in the vertical plane of the chamber. 
This configuration allows the ice samples in either substrate to face the electron or vacuum ultraviolet (VUV) beams used for the energetic processing (see Sections \ref{sec:electron} and \ref{sec:uv}, respectively), as well as the infrared (IR) spectrometer (Sect. \ref{sec:ir}), quadrupole mass spectrometer (QMS, Sect. \ref{sec:tpd}), and reflectron time-of-flight mass spectrometer (Re-ToF MS, Sect. \ref{sec:tof}) that are used to monitor the evolution of the ice samples during the experimental simulations, as sketched in Fig. \ref{fig:ST_sketch}. 

\begin{figure*}
    \centering
    \plottwo{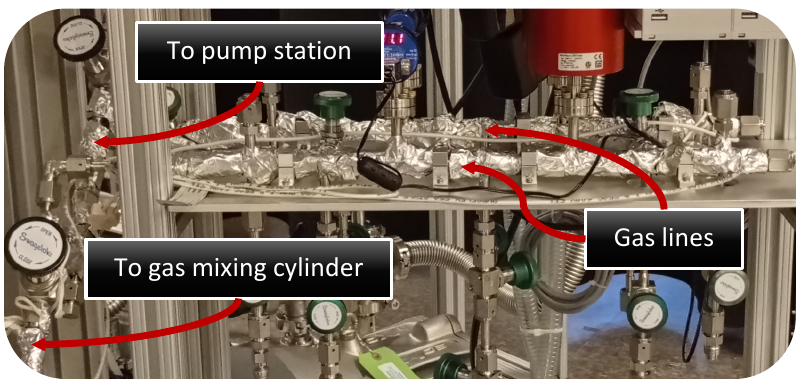}{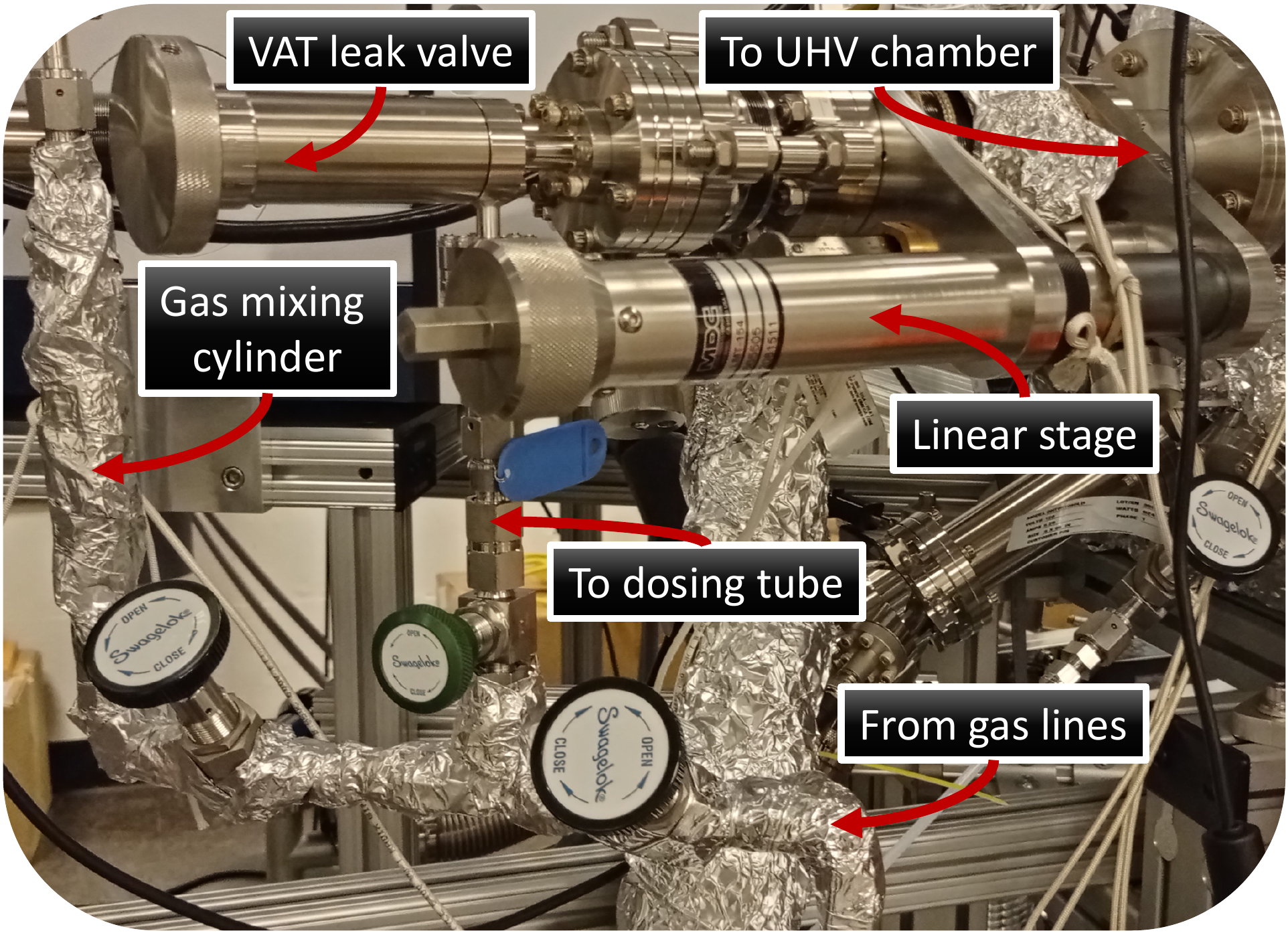}
    \caption{Gas handling assembly consisting of two interconnected lines with ports for the introduction of different species 
    (\textit{left panel}) and a 300 ml gas mixing cylinder (\textit{right panel}). The right panel also shows the doser assembly, composed of a VAT variable leak valve and a 10 mm diameter dosing tube mounted on a linear stage.}
    \label{fig:ST_gasline}
\end{figure*}

The deposition of the ice samples in SPACE TIGER is performed by exposing one of the substrates on the sample holder to a constant flow of a gas mixture with the desired composition from an independently pumped stainless steel gas line assembly. 
This gas handling assembly consists of two interconnected identical gas lines (that can be sectioned off if needed), which in turn are connected to a 300 ml mixing cylinder where the gas mixture is made and stored, as shown in Fig. \ref{fig:ST_gasline}. 
The gas line assembly is pumped by  
an Edwards TIC pumping station with a NEXT85D turbomolecular pump and a nXDs 6i scroll pump 
down to a $\sim$10$^{-7}$ Torr base pressure. 
The pressure of the two lines is measured with two Baratron pressure gauges (Pfeiffer CMR371 and CMR372), while the pressure of the mixing cylinder is monitored with a PCR 280 Baratron pressure gauge. 
The two lines contain ports for small gas cylinders, stainless steel canisters, and glass flasks used to introduce different species through on/off valves. 
Opening an isolation valve allows the gas mixtures to flow through a VAT variable leak valve into the 10 mm diameter dosing tube mounted on a MDC LMT-154 linear stage through which they are dosed into the chamber at very close proximity ($\sim$2 mm) from the substrate. 
We thus assume a 10 mm diameter (the same diameter as the dosing tube) for the ice samples deposited on both the copper and CsI substrates. 
The dosing tube is isolated by a bellows sealed metal valve (SS-4H-VCR). 
The dosing process is terminated by quickly evacuating the gas line. 
The doser assembly is presented in the right panel of Fig. \ref{fig:ST_gasline}. 

For a reproducible dosing rate, a laser drilled 7x7, 4 $\mu$m diameter pinhole array (Lenox Laser, custom made) is mounted on a VCR flanged enclosure behind the VAT variable leak valve of the doser assembly, which is set to the fully open position during the dosing process. 
A rough estimation of the deposited ice thickness can be calculated assuming that all molecules entering the chamber are deposited onto the substrate with a sticking coefficient of unity. The ice column density at the end of the dosing process 
would thus be equal to the total number of molecules that have entered the chamber divided by the surface area of the ice on the substrate. 
If the pressure in the gas line assembly is lower than 20 Torr, the mean free path of the particles is larger than the size of the pinholes, and the total number of molecules entering the chamber can be calculated as the product of the effusive flow rate $Q_{effusion}$ and the dosing time $t$. 
The effusive flow rate is: 
\begin{equation}
    Q_{effusion}=\frac{P A}{\sqrt{2 \pi m k_B T}},
\end{equation}
where $P$ and $T$ are the pressure and the temperature in the gas line assembly, $k_B$ is the Boltzmann constant, $m$ is the species molecular mass, and $A$ is the total area of the pinhole array. 
Assuming that the molecules in the gas line assembly are at room temperature and that the deposited ice is 10 mm diameter (see above), 
the ice column density at the end of the deposition process would be
\begin{equation}
    N \approx \frac{16  P  t}{\sqrt{M}},
    \label{eq:N_dosing}
\end{equation}
\noindent where $N$ is the number of monolayers (1 ML = 10$^{15}$ molecules cm$^{-2}$), $P$ is the pressure in the gas mixing cylinder during dosing (in Torr), $t$ is the dosing time in seconds, and $M$ is the molar mass of the dosed species. 
In the case of ice mixtures, the column densities of the different ice components could be estimated with Eq. \ref{eq:N_dosing} using the corresponding partial pressures. 
The ice column density estimated with Eq. \ref{eq:N_dosing} should be considered an upper limit, since it assumes that all molecules entering the chamber are deposited onto the substrate with a sticking coefficient of unity, as explained above. 

The column density of the different ice species can be empirically calculated from the integrated absorbance of the corresponding IR features when the ice samples are deposited onto the CsI substrate, as explained in Sect. \ref{sec:ir}. 
In those cases, we have found that the ice column densities calculated from the IR absorbance are 2$-$3 times lower than the upper limits estimated with Eq. \ref{eq:N_dosing}. 
This difference could be larger for ice components whose behavior diverge from the ideal gas law (for example in the case of semi-volatile species). 
Deviations from the theoretical estimation are probably due to the combination of multiple factors. It is likely that not all the molecules entering the chamber get to the copper or CsI substrates, as a fraction will be probably pumped out before this happens (even for a short distance between the end of the dosing tube and the substrate).  In addition, the molecule sticking coefficients may be lower than unity, especially with higher flow rates and higher substrate temperatures.

In this work, we used $c$-C$_6$H$_6$ (degassed, $>$99.9\% purity, Sigma-Aldrich), CH$_3$CN (liquid, 99.8\%, Sigma-Aldrich), H$_2$O (liquid, deionized, Sigma-Aldrich), and CO (gas, 99.95\%, Aldrich) as the initial components of the gas mixture.  The ice samples were grown on the copper substrate, and the temperature was set to $\sim$4$-$10 K during dosing and irradiation. 

\subsection{Electron irradiation of the ice samples}\label{sec:electron}

Electron irradiation of the deposited ice samples is carried out using a ELG-2/EGPS-1022E low energy electron source (Kimball Physics) with variable energy (1eV to 2keV) , beam current (1nA to 10uA), spot size (0.5mm to 5mm), and fast pulsing (20 ns to 100 $\mu$s) capabilities. 
The electron beam current is measured with a Faraday cup placed at the tip of the electron source. 
%
%
For the experiments presented in this work, we fixed the energy of the electrons to 2 keV, and the average electron beam current to 50$-$52 nA (relative uncertainty of 10\%). 
The sample irradiation time was set to 56$-$60 min, so that the ice samples were processed with a total of 1.10$-$1.15 $\times$ 10$^{15}$ electrons, corresponding to a total irradiated energy of 2.2$-$2.3 $\times$ 10$^{18}$ eV, or a total energy fluence of 2.8$-$2.9 $\times$ 10$^{18}$ eV cm$^{-2}$ over the 10 mm diameter ice surface. 
%
The sample irradiation time was selected to match the estimated energy fluence directly deposited by the cosmic rays into the interstellar ices during the typical lifetime of a dense cloud 
\citep[$\sim$ 10$^6$ yr, see][and ref. therein]{jones11}. 

We note that the 2 keV electron penetration depth in the ice samples was expected to be lower than the total ice thickness, according to the electron penetration depths reported for ice samples with a similar density in \citet{martin20}. 

\begin{figure*}
    \centering
    \includegraphics[width=10cm]{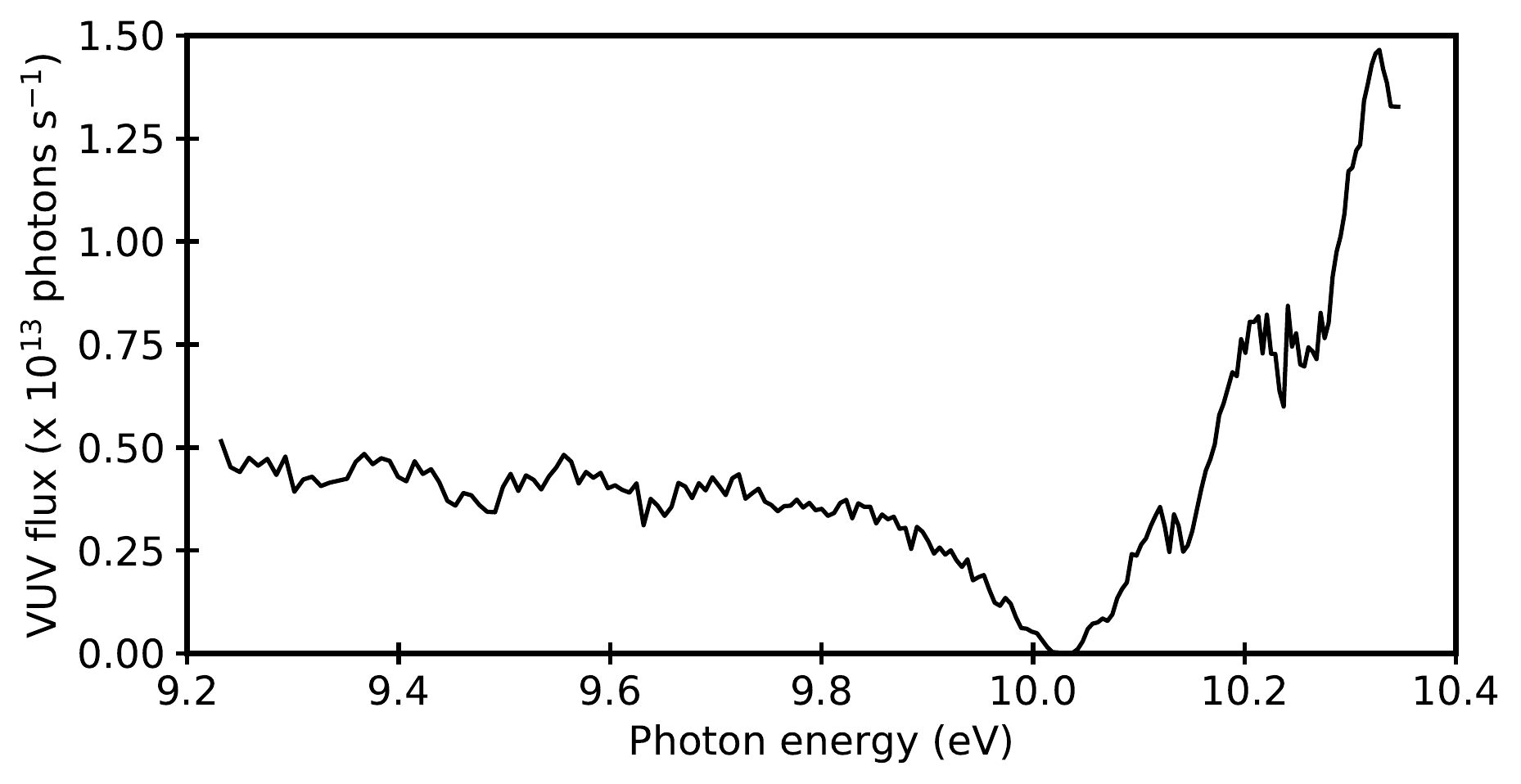}
    \caption{VUV photon flux as a function of the photon energy in SPACE TIGER}
    \label{fig:vuv}
\end{figure*}

\subsection{VUV photoprocessing of the ice samples}\label{sec:uv}

The ice samples in SPACE TIGER can also be processed with vacuum ultraviolet (VUV) photons generated by a tunable laser system. 
VUV photons are also needed for the single photon ionization (SPI) required for the Re-ToF mass spectrometry detection of desorbing ice molcules (Sect. \ref{sec:tof}).  
%
VUV photon pulses in SPACE TIGER are produced \textit{via} two-photon resonance-enhanced difference frequency nonlinear mixing (also known as resonance four-wave frequency mixing), 
using krypton (Kr) as the nonlinear mixing medium \citep{marangos90}. 
In this frequency mixing process, the resonant state 4s$^2$4p$^5$($^2$P$_{1/2}$)5p$^1$ (12.258 eV) in Kr is populated by two 202.316 nm UV photons. 
Then, the energy of a third photon in the visible range (410 nm $-$ 650 nm) is subtracted to produce a VUV photon in the 120 nm $-$ 135 nm range, corresponding to energies between 9.2 eV and 10.4 eV. 

In SPACE TIGER, 202.316 nm laser pulses are generated by a NarrowScan dye laser (Radiant Dyes Laser \& Accessories GmbH), pumped by the second harmonic of a LAB-190 Nd:YAG laser from SpectraPhysics (component 7a in Fig. \ref{fig:ST_picture}) at 532 nm, with an energy of $\sim$170 mJ/pulse. 
An electronic pulse is supplied to the Nd:YAG laser by a digital delay pulse generator Series 9528 (Quantum Composers) with four 50 Ohm outputs and four high-Z outputs, leading to a 10 ns pulse duration and a pulse frequency of 30 Hz. 
The dye laser is equipped with a Bethune cell amplifier where the mixture of Rhodamine 610 and Rhodamine 640 dyes diluted in ethanol is used to produce the fundamental radiation at 606.948 nm. 
The 606.948 nm radiation is frequency doubled (303.474 nm) and tripled (202.316 nm) passing through the frequency doubling and tripling crystals integrated inside the laser housing. 
The crystals are kept at a temperature of 38 degrees Celsius. 
The three resulting wavelengths (606.948, 303.474, and 202.316 nm) are dispersed by a Pellin-Broca prism. 
The wavelengths emitted by the dye laser can be measured with a WaveMaster wavelength meter (Coherent).  
The 202.316 nm output power is $\sim$1.7 mJ. 

At the same time, tunable visible laser pulses are generated by a primoScan optical parametric oscillator (OPO, GWU-Lasertechnik) pumped by the third harmonic of a PRO-250 Nd:YAG laser also from SpectraPhysics (component 7b in Fig. \ref{fig:ST_picture}) at 355 nm, with an energy of $\sim$320 mJ. 
This OPO can produce laser pulses in the UV, visible, and infrared (IR) range (between 250 nm and 2000 nm). 
As mentioned above, visible pulses between 410 nm and 650 nm are used for VUV photon production in the 120 nm $-$ 135 nm range. 
The wavelengths emitted by the OPO are tuned through the corresponding software GWU Motion Master. 

The 202.316 nm pulse from the dye laser is combined with the visible pulse from the OPO on a dichroic mirror (CVI, TLM1-200-45-2037), and subsequently focused with an F = 300 mm lens to the overlapping spots inside a cell filled with $\sim$2.8 Torr of Kr where the resonance four-wave frequency mixing process explained above takes place. 
This cell is evacuated by a Leybold Turbolab 90i pump station (90 ls$^{-1}$). 

When the wavelength of the generated VUV photons corresponds to the Ly-$\alpha$ transition (121.567 nm, 10.20 eV), the 202.316 nm pulse from the dye laser can be combined with the dye laser fundamental radiation at 606.948 nm, instead of using the visible laser pulse from the OPO. 
In that case, $\sim$20 Torr of a 3:1 Ar:Kr mixture is used as the mixing medium,  
since the addition of argon improves phase-matching conditions for the difference frequency mixing process 
at this particular wavelength \citep{marangos90}. 

The produced VUV photons are guided into the chamber by a guiding arm pumped by a Leybold MAG W300 iP pump with a pumping capcity of 300 ls$^{-1}$, backed by a Pfeiffer MVP 015-4 diaphragm pump. 
The VUV radiation produced at the focal spot is refocused onto a 1.5 mm diameter aperture inside the arm guiding the VUV beam with an off-axis LiF lens (plano-convex, 70 mm ROC, Crystan Ltd.), in an arrangement similar to the one used by \citet{hanna09}.
The aperture blocks the 410$-$650 nm and 202.316 nm wavelengths, which are dispersed differently from the VUV by the off-axis LiF lens. 
The VUV pulse can be collected with a Hamamatsu S10043 photodiode located at the opposite port of the chamber and measured with a RTO2014 oscilloscope (Rhode \& Schwarz). 
The photodiode has been calibrated at Physikalisch-Technische Bundesanstalt (PTB) in Berlin (Germany). 
The measured VUV photon flux produced when using Kr as the mixing medium is presented in Fig. \ref{fig:vuv}. 
We note that the drop in the VUV photon flux observed around 10.03 eV in Fig. \ref{fig:vuv} corresponds to the electronic excitation energy from the ground state to the 4s$^2$4p$^5$($^2$P$_{3/2}$)5s excited state in Krypton. The VUV photons at this wavelength were re-absorbed by the gas-phase Krypton atoms inside the gas cell. 
The addition of Ar to the mixing medium increases the VUV photon flux at the Ly-$\alpha$ wavelength by a factor of $\sim$2 with respect to that presented in Fig. \ref{fig:vuv}.

\subsection{IR ice spectroscopy of the ice samples}\label{sec:ir}

The evolution of the ice samples during irradiation can be monitored by means of a Bruker 70v Fourier transform infrared (FTIR) spectrometer equipped with two liquid-nitrogen-cooled MCT detectors. 
Both the spectrometer and the MCT detector chambers are evacuated to 2 mbar to avoid atmospheric IR features in the ice spectrum using two Edwards scroll pump nXDs 15i (4 l s$^{-1}$).
The FTIR spectrometer can be operated in either the transmission or reflection-absorption mode, depending on the substrate used to grow the ice sample (CsI or copper, respectively). 

The species ice column densities can be calculated from the IR spectra using the equation  

\begin{equation}
N=\frac{1}{A}\int_{band}{\tau_{\nu} \ d\nu}
\label{eqn}
\end{equation}

\noindent (assuming a normal incidence of the IR beam), where $N$ is the ice column density in molecules cm$^{-2}$, $\tau_{\nu}$ the optical depth of a particular absorption band (2.3 times the absorbance), and $A$ the band strength in cm molecule$^{-1}$. 
Most of the reported IR band strengths in the literature correspond to IR spectra collected in transmission mode. 
Setup-specific band strengths for reflection-absorption IR spectroscopy can be calculated. 
However, the IR absorbance measured in reflection-absorption mode shows a non-linear behavior with the species column densities when the ice thickness is above a certain threshold \citep[see, e.g.,][]{oberg09}. This is the case of the ice samples listed in Table \ref{tab:exp}.
In those cases, Eq. \ref{eq:N_dosing} can be used to estimate approximate ice thicknesses (Sect. \ref{sec:dosing}). 
While this is accurate enough for the exploratory study presented in this work, different techniques are needed if the aim is to provide quantitative reaction data. 
In the latter case, there are calibration schemes available to improve ice thickness measurements of ice samples deposited onto the copper substrate, or ice samples that are IR-inactive \citep[see, e.g.,][]{martin20,tajana22}. 

In this work, reflection-absorption IR spectra were collected with a resolution of 1 cm$^{-1}$ in the 5000 $-$ 600 cm$^{-1}$ range, and averaged over 128 interferograms. When operated in reflection-absorption mode, the IR beam has an incidence angle of 20$^\circ$ with respect to the substrate surface. 
%

\begin{deluxetable*}{ccccccc}
\tablecaption{Summary of experiments simulating the energetic processing of ice samples containing $c$-C$_6$H$_6$ and CH$_3$CN}
\label{tab:exp}
\tablehead{\colhead{Exp.} & \colhead{Ice composition} & \colhead{Comp. ratio$^b$} & \colhead{Ice thickness$^b$} & \colhead{Irradiation} & \colhead{Irradated energy} & \colhead{Ionizing VUV photon flux}\\
&&&(ML)&&\colhead{($\times$10$^{18}$ eV)} & \colhead{($\times$10$^{11}$ photons/pulse)}}
\startdata
1 & $c$-C$_6$H$_6$$^a$ & - & 1300 & 2 keV e$^-$ & 2.08 & 5.77\\
2 & CH$_3$CN & - & 1345 & 2 keV e$^-$ & 2.25 & 5.13\\
\hline
3 & $c$-C$_6$H$_6$:CH$_3$CN & 1:1 & 1130 & 2 keV e$^-$ & 2.20 & 5.18\\
4 & $c$-C$_6$H$_6$:CH$_3$CN & 1:1 & 1130 & 2 keV e$^-$ & 2.18 & 5.37\\
5 & $c$-C$_6$H$_6$:CH$_3$CN & 1:1 & 1130 & 2 keV e$^-$ & 2.25 & 2.37 / 1.56$^c$\\
6 & $c$-C$_6$H$_6$:CH$_3$CN & 1:1 & 1130 & Ly-$\alpha$ & 1.39 & 4.74\\
\hline
7 & H$_2$O:$c$-C$_6$H$_6$:CH$_3$CN & 10:1:1 & 1880 & 2 keV e$^-$ & 2.30 & 5.13\\
8 & H$_2$O:$c$-C$_6$H$_6$:CH$_3$CN & 10:1:1 & 1960 & Ly-$\alpha$ & 1.36 & 5.43\\
\hline
9 & CO:$c$-C$_6$H$_6$:CH$_3$CN & 10:1:1 & 1660 & 2 keV e$^-$ & 2.30 & 5.43\\
10 & CO:$c$-C$_6$H$_6$:CH$_3$CN & 10:1:1 & 1660 & Ly-$\alpha$ & 1.44 & 6.05\\
\enddata
\tablecomments{
Ice samples were grown and irradiated at T $\sim$ 4$-$10 K. 
$^a$This experiment was performed two times. 
In the first run, the dosing and irradiation temperature was 50 K, and no IR spectra were collected.
In the second run, the dosing and irradiation temperature was 10 K. 
Unfortunately, experimental complications prevented us from using the Re-ToF MS during the second run. 
Therefore, the Re-ToF MS data shown in Figures \ref{fig:re-tof-ms} and \ref{fig:ms} correspond to the first run.
$^b$Ice composition ratio and thickness were estimated using Eq. \ref{eq:N_dosing} as a first approximation. 
$^c$These two values correspond to the VUV photon flux of the two ionizing VUV photon energies (10.2 eV and 9.7 eV) that were alternatively used during the TPD of the irradiated ice sample in this experiment.}
\end{deluxetable*}

\subsection{Sample temperature control and Temperature Programmed Desorption}\label{sec:tpd}

The temperature of the ice samples is controlled and monitored by a LakeShore 336 temperature controller using a calibrated Si diode sensor with a 2 K estimated accuracy and a 0.1 K relative uncertainty.
The ice temperature can be set to any value between $\sim$4 K and room temperature thanks to the combination of the closed-cycle He cryostat and a 50 W silicon nitride cartridge heater rod (Bach Resistor Ceramics) pressed into a matching grove in the OFHC copper sample holder, behind the copper and CsI substrates. 
Temperature programmed desorption (TPD) of the ice samples is usually carried out at the end of any experiment. 
During TPD, a constant heating rate (2 K min$^{-1}$ in the case of the experiments presented in this work) is applied to the substrate until the complete sublimation of the ices is achieved. 
The desorbing molecules during TPD are detected with either the QMG 220M1 QMS (Pfeiffer, mass range 1$-$100 amu, resolution of 0.5 amu) or, as it was the case for the experiments presented in this work, the reflectron time-of-flight mass spectrometer (Sect. \ref{sec:tof}). 

\subsection{Reflectron Time-of-Flight mass spectrometry of the desorbing molecules}\label{sec:tof}

The desorbing molecules can be detected \textit{via} single-photon ionization (SPI) reflectron time-of-flight (Re-ToF) mass spectrometry, using a D-850 Jordan angular Re-ToF mass spectrometer (MS) designed for the use with grounded ice samples (i.e., ice samples deposited on the copper substrate). 
This detection technique is more sensitive than the electron-impact ionization mass spectrometry, and allows to discriminate between different isomers based on their ionization energies \citep[for a detailed description of the technique see]{boesl91}. 
The Re-ToF MS is equipped with a D-803i power supply, and an additional power supply for the extraction plate.  
It is attached to the UHV chamber through two HRW50CA1SSC1 blocks, sliding on a HRW50-760L rail (THK America) \textit{via} custom built aluminium clamps. 
This allows the Re-ToF MS to move closer to the sample before measurements. 
The movement is driven by a McAllister 4 inch motorized translation through an 8 inch CF flange BLT86-04-M0. 
The flight tube of the Re-ToF MS is evacuated by a Leybold MAG W600 iP pump, with a pumping capacity of 600 l s$^{-1}$, also backed by the same magnetically levitated Leybold MAG W300 iP pump (300 l s$^{-1}$) 
and the Edwards scroll pump nXDs 15i (4 l s$^{-1}$) that are used to back the magnetically levitated turbomolecular pump of the UHV chamber. 

In order to be detected by the Re-ToF MS, the desorbing molecules must be first ionized by a VUV photon from the tunable laser system described in Sect. \ref{sec:uv}. 
When the ionizing photon energy is close to the photoionization threshold of the detected species, fragmentation inside the MS is minimized, which is in contrast to the fragmentation-prone electron impact ionization used in the QMS. 
The selection of a working ionizing photon energy for the experiments presented in this work is explained in Sect. \ref{sec:pie}. 
The ions are subsequently detected using a multichannel plate with a dual chevron configuration. 
The detector reading is synchronized with the ion pulse generated by the tunable VUV laser system. 
A 30 Hz pulse is generated by the digital delay pulse generator Series 9528 from Quantum Composers presented in Sect. \ref{sec:uv}. 
The detected signals are amplified using a fast preamplifier (Ortec 9305) and shaped with a 100 MHz discriminator. 
Finally, the Re-ToF spectra are recorded with a personal-computer-based multichannel scaler (FAST ComTec, P7888-1 E). 

The MS can detect ions with a mass-to-charge ($m/z$) ratio of up to 555 amu. 
The $m/z$ ratio of the detected ions is determined by their residence times in the Re-ToF MS flight tube, with longer residence times corresponding to heavier ions. 
The mass calibration of the Re-ToF MS was performed with the following molecules: ammonia (NH$_3$), hydrogen sulfide (H$_2$S), allene (C$_3$H$_4$), propene (C$_3$H$_6$), 1,3-butadiene (C$_4$H$_6$), n-butane (C$_4$H$_{10}$), n-pentane (C$_5$H$_{12}$), n-hexane (C$_6$H$_{14}$), vinyl bromide (C$_2$H$_3$Br), and iodotrifluoromethane (CF$_3$I). 

A 4 ns bin width, i.e. a 4 ns arrival delay to the detector, is accounted for every measured $m/z$. 
The theoretical mass resolution varies with the $m/z$ value, and ranges from $\sim$0.01 amu at $m/z$ = 1 to $\sim$0.2 amu at $m/z$ = 500. 
However, the detection of ions with mass $m$ corresponding to a given species leads to measured signals over a small range of $m/z$ values, 
suggesting that the arrival delay to the detector is larger than 4 ns for ions of the same species.  
We observe that the ion arrival times follow an approximately normal distribution. 
We therefore fit each detected $m/z$ peak with a Gaussian function in order to obtain the integrated ion signal corresponding to a particular species.  
We find a typical full-width-half-maximum (FWHM) around 0.15 amu at $m/z$$\sim$100, which implies a lower effective mass resolution, especially for species with low $m/z$. 
In addition, we find that the $m/z$ position of the peak corresponding to a given $m/z$ fragment may vary from one experiment to another by up to 0.7 amu. 
The integrated ion signal as a function of the temperature during the TPD of the processed ice samples represents the so-called TPD curve for a particular species.

\section{Results}\label{sec:results}

As outlined in the Introduction, the main goal of this work is to address the formation of $c$-C$_6$H$_5$CN upon 2 keV electron and Ly-$\alpha$ photon irradiation of ice samples containing $c$-C$_6$H$_6$ and CH$_3$CN, where CH$_3$CN acts as a source of CN radicals. 
The presence of $c$-C$_6$H$_5$CN molecules in the irradiated ice samples was explored by means of reflection-absorption infrared spectroscopy (Sect. \ref{sec:ir}) and reflectron time-of-flight mass spectrometry upon thermal desorption of the irradiated ice samples (Sect. \ref{sec:tof}). 
The latter required a previous measurement of the $c$-C$_6$H$_5$CN photoionization efficiency (PIE) in the gas phase, which is presented in Sect. \ref{sec:bn} of the Appendix. 
The energetic processing of binary $c$-C$_6$H$_6$:CH$_3$CN
1:1 ice mixtures is presented in Sections \ref{sec:exp_binary_e}$-$\ref{sec:exp_binary_ph}, and compared to the 2 keV electron irradiation of pure $c$-C$_6$H$_6$ and CH$_3$CN ices. 
Sect. \ref{sec:exp_ternary} presents the experiments with more realistic ice samples in which $c$-C$_6$H$_6$ and CH$_3$CN molecules were embedded in a H$_2$O or CO ice matrix. The experiments presented in Sections \ref{sec:exp_binary_e}$-$\ref{sec:exp_ternary} are listed in Table \ref{tab:exp}. 

\begin{figure*}
    \centering
    \includegraphics[width=11cm]{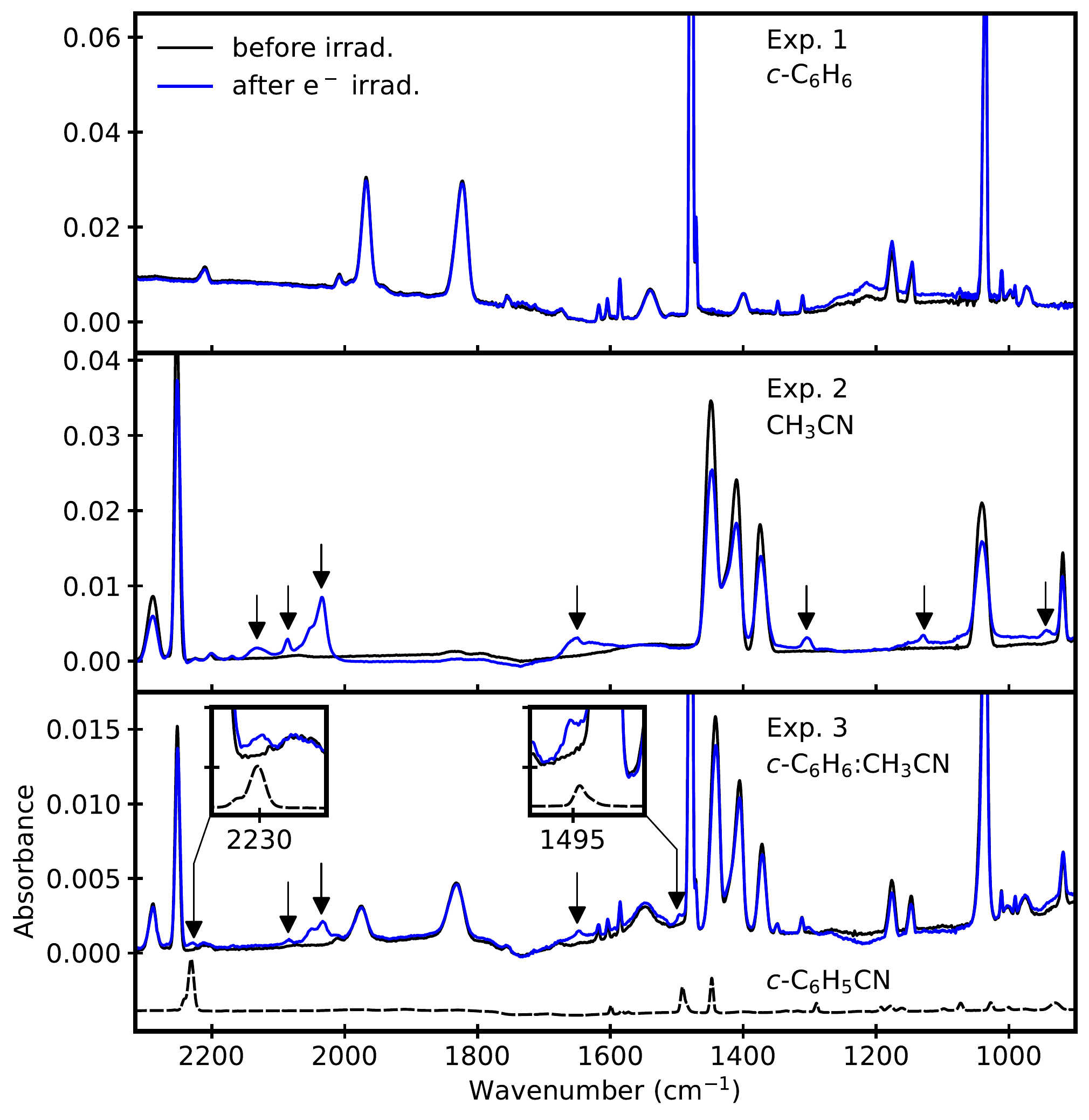}
    \caption{IR spectra collected in reflection-absorption mode before (black) and after (blue) irradiation of an incident energy of $\sim$2.2 $\times$ 10$^{18}$ eV with 2 keV electrons of $c$-C$_6$H$_6$ and CH$_3$CN ice samples (top and middle panels, Experiments 1 and 2 in Table \ref{tab:exp}), and a 1:1 $c$-C$_6$H$_6$:CH$_3$CN ice mixture (bottom panel, Exp. 3). 
    The IR spectrum of a pure $c$-C$_6$H$_5$CN ice (black dashed line) is also shown as a reference in the bottom panel.
    Arrows in the middle and bottom panels indicate new IR features detected after processing of the ice samples. 
    The two insets in the bottom panel show a zoom-in of the two IR features assigned to $c$-C$_6$H$_5$CN, along with the corresponding IR bands from the pure $c$-C$_6$H$_5$CN ice spectrum (scaled down to a comparable absorbance).
    }
    \label{fig:ir}
\end{figure*}

\begin{figure*}
    \centering
    \gridline{
    \fig{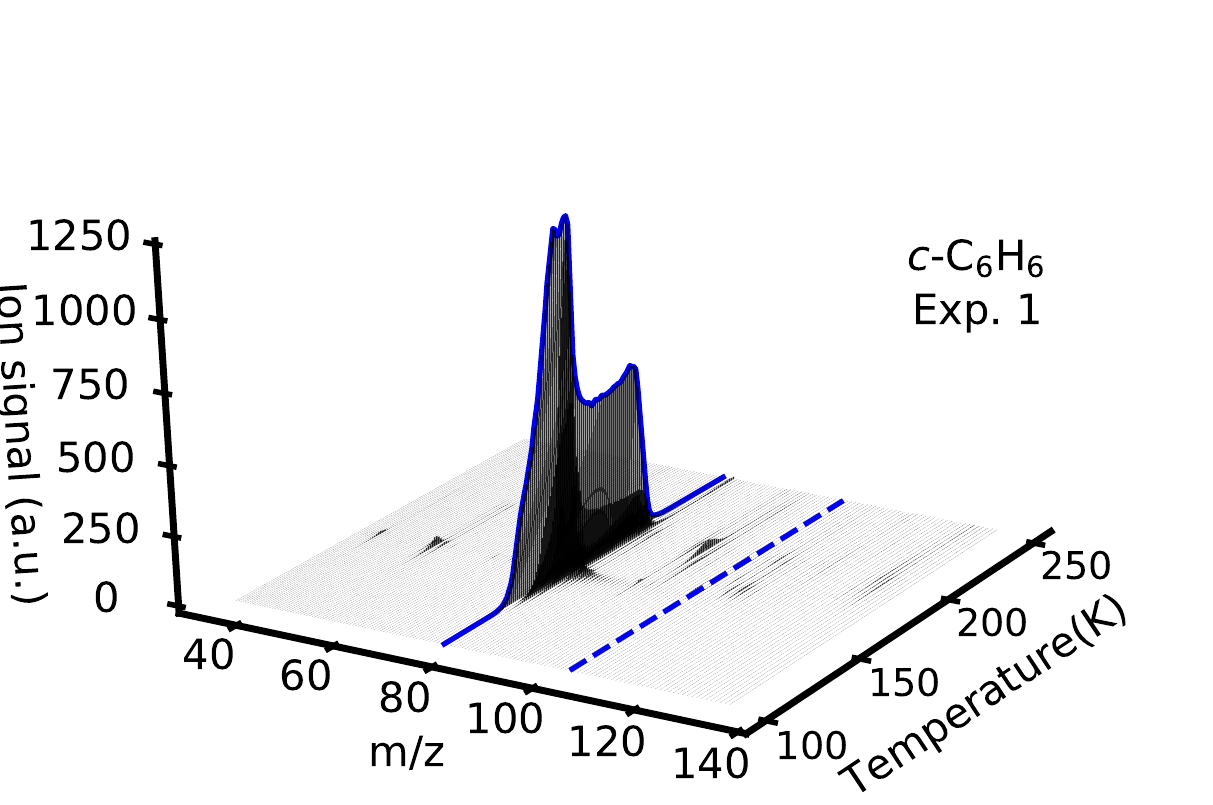}{0.45\textwidth}{}
    \fig{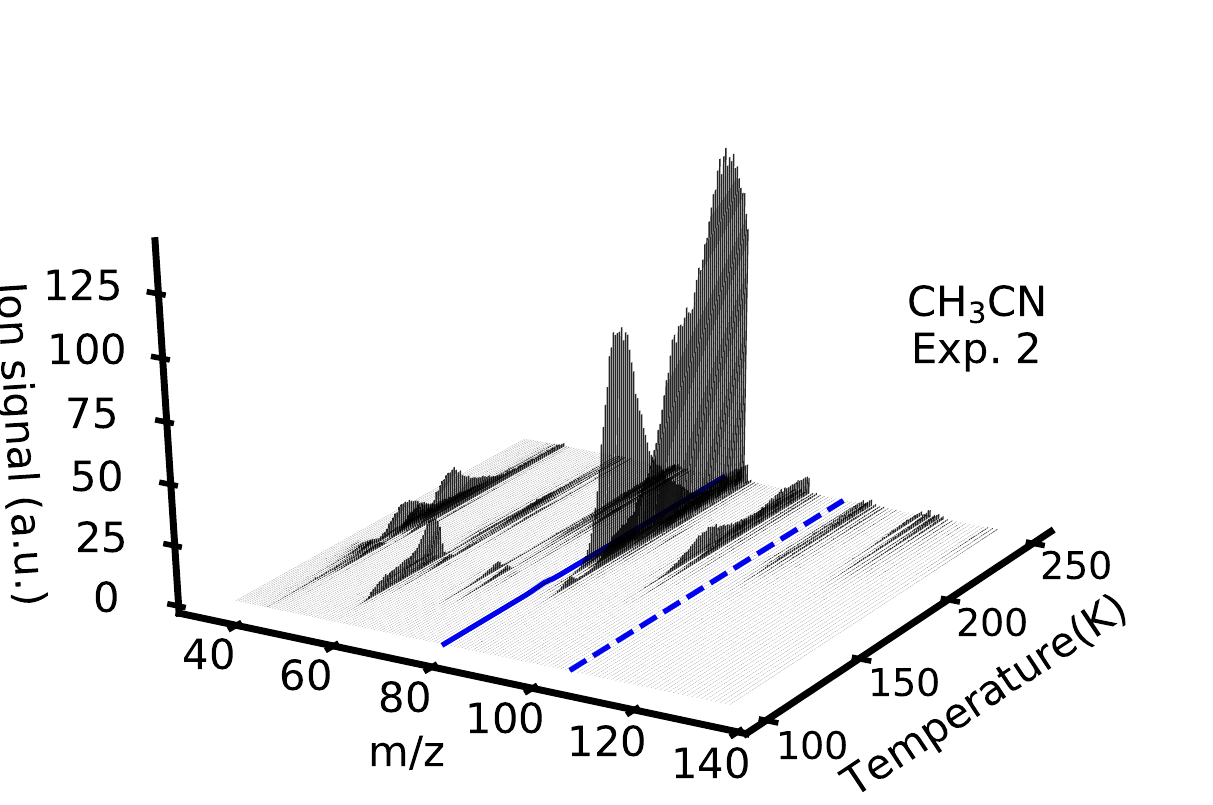}{0.45\textwidth}{}
    }
    \vspace{-11mm}
    \gridline{ 
    \fig{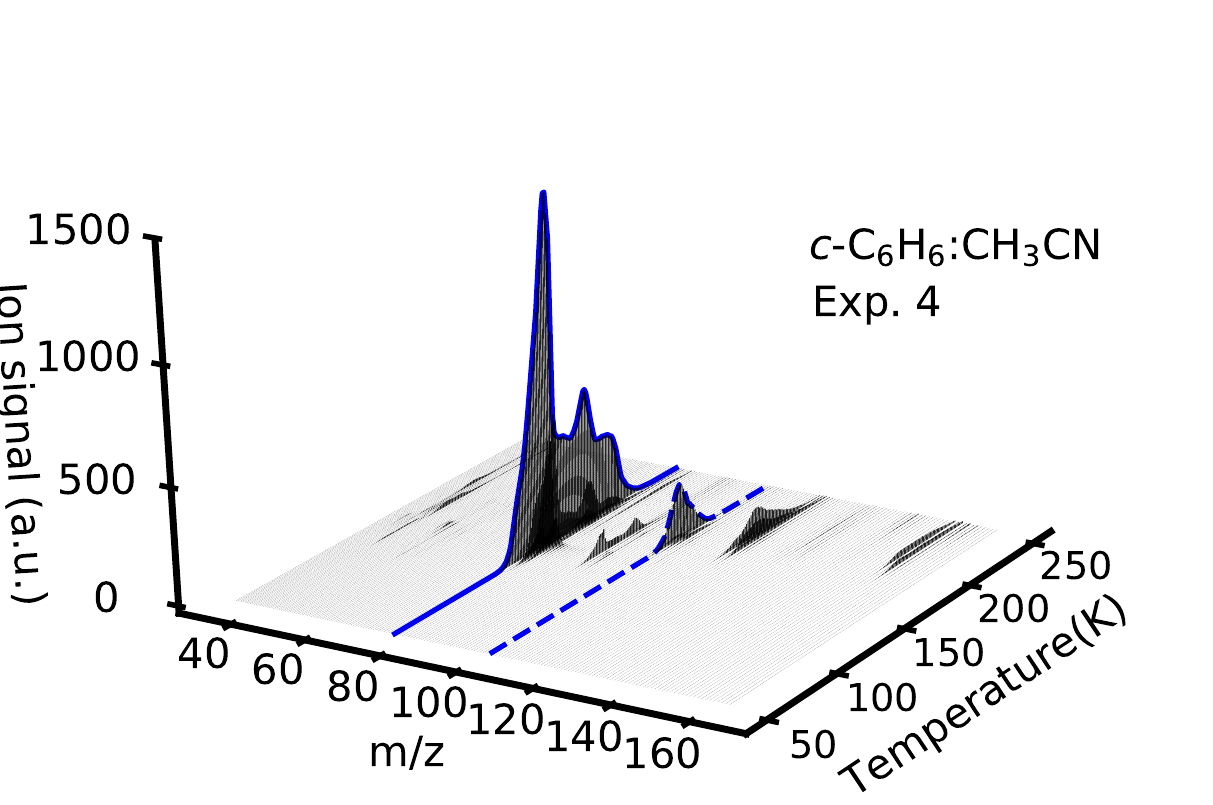}{0.45\textwidth}{}
    }
    \caption{Re-ToF MS data collected during the TPD of the ice samples after 2 keV electron irradiation of pure $c$-C$_6$H$_6$ and CH$_3$CN ices (Experiments 1 and 2, top left and right panels, respectively), and a $c$-C$_6$H$_6$:CH$_3$CN ice mixture (Exp. 4, bottom panel).  
    The measured ion signals are represented as a function of the mass-to-charge ratio ($m/z$) and the temperature. 
    For clarity, the ion signals were added every 17 $m/z$ values (corresponding to $\sim$0.5 amu at $m/z$ = 35 and $\sim$1 amu at $m/z$ = 165) and assigned to the middle $m/z$ value. 
    The position of $m/z$ $\approx$ 78 and $m/z$ $\approx$ 103 (corresponding to $c$-C$_6$H$_6$ and $c$-C$_6$H$_5$CN) is highlighted with solid and dashed blue lines, respectively. }
    \label{fig:re-tof-ms}
\end{figure*}

\subsection{2 keV electron irradiation of a 1:1 $c$-C$_6$H$_6$:CH$_3$CN ice mixture}\label{sec:exp_binary_e}

%
Experiments 3$-$5 in Table \ref{tab:exp} consisted in $\sim$2.2 $\times$ 10$^{18}$ eV irradiation with 2 keV electrons of a 1:1 $c$-C$_6$H$_6$:CH$_3$CN ice sample. 
The bottom panel of Fig. \ref{fig:ir} shows the ice IR spectrum in the 2315$-$900 cm$^{-1}$ range collected in the reflection-absorption mode before and after the 2 keV electron irradiation in Exp. 3 (the results were similar in Exp. 5, while no IR spectra were collected for the ice sample in Exp. 4). 
For comparison, the top and middle panels of Fig. \ref{fig:ir} show the IR spectra in the same range before and after 
2 keV electron irradiation of $c$-C$_6$H$_6$ and CH$_3$CN ice samples, respectively (Experiments 1 and 2 in Table \ref{tab:exp}). 
Five new IR features were detected after irradiation of the $c$-C$_6$H$_6$:CH$_3$CN ice mixture at, approximately, 2230 cm$^{-1}$, 2085 cm$^{-1}$, 2035 cm$^{-1}$, 1650 cm$^{-1}$, and 1495 cm$^{-1}$ (Fig. \ref{fig:ir}, bottom panel). 
The features at 2085 cm$^{-1}$, 2035 cm$^{-1}$, and 1650 cm$^{-1}$ were also detected after processing of the CH$_3$CN ice sample (middle panel of Fig. \ref{fig:ir}),  
and thus probably corresponded to ice chemistry products derived solely from CH$_3$CN whose assignment was not pursued.
The features at 2230 cm$^{-1}$ and 1495 cm$^{-1}$, on the other hand, 
were only detected after irradiation of an ice sample containing both $c$-C$_6$H$_6$ and CH$_3$CN.  
These two features could correspond to $c$-C$_6$H$_5$CN based on the observed IR bands at 2231 cm$^{-1}$ and 1491 cm$^{-1}$ in a pure $c$-C$_6$H$_5$CN ice spectrum, also measured as part of this work and shown in Fig. \ref{fig:ir} as a reference. 
Other $c$-C$_6$H$_5$CN ice IR features are either weaker or overlap with bands corresponding to CH$_3$CN 
(1447 cm$^{-1}$, 763 cm$^{-1}$) or $c$-C$_6$H$_6$ (690 cm$^{-1}$) (the latter are not shown in Fig. \ref{fig:ir}).

After irradiation, the ice samples in Experiments 1$-$5 were warmed up to room temperature, and  
the desorbing molecules were subsequently detected by the Re-ToF MS using Ly-$\alpha$ photons as the ionizing radiation (see Sect. \ref{sec:bn} of the Appendix). 
Figure \ref{fig:re-tof-ms} shows the measured ion signals, 
as a function of the mass-to-charge ratio ($m/z$) and the temperature, detected after irradiation of $c$-C$_6$H$_6$ and CH$_3$CN ices (Experiments 1 and 2, top left and right panels, respectively), and a $c$-C$_6$H$_6$:CH$_3$CN ice mixture (Exp. 4, bottom panel - the results were similar in Experiments 3, 4 and 5). 
We note that the Re-ToF MS data presented for the irradiated $c$-C$_6$H$_6$ ice sample correspond to the first of the two runs carried out for Exp. 1 where the dosing and irradiation temperature was 50 K, as indicated in Table \ref{tab:exp}. 
No significant chemistry was detected during electron irradiation of a $c$-C$_6$H$_6$ ice sample at 10 K or 50 K.
%
The temperature-dependent mass spectrum of the processed $c$-C$_6$H$_6$ ice sample was dominated by a peak at $m/z$ $\approx$ 78 (Fig. \ref{fig:re-tof-ms}, top left panel), 
corresponding to the unreacted $c$-C$_6$H$_6$ molecules, along with less intense peaks around $m/z$ $\approx$ 79 and $m/z$ $\approx$ 80 that correspond to $c$-C$_6$H$_6$ isotopologs where one or two C atoms have been substituted by $^{13}$C. 
The most intense peak in the mass spectrum of the irradiated CH$_3$CN ice sample was located at $m/z$ $\approx$ 82 (Fig. \ref{fig:re-tof-ms}, top right panel). 
This feature is an order of magnitude weaker than the $m/z$ $\approx$ 78 peak in Exp. 1, and we speculate that it might originate from clusters formed by unreacted CH$_3$CN molecules. 
CH$_3$CN molecules have an ionization energy above 10.20 eV, and those not forming clusters were barely detected in any of the experiments. 
In addition to the $m/z$ $\approx$ 78 and $m/z$ $\approx$ 82 peaks, additional peaks with higher $m/z$ were detected after irradiation of a $c$-C$_6$H$_6$:CH$_3$CN ice mixture (Fig. \ref{fig:re-tof-ms}, bottom panel), including a peak located at $m/z$ $\approx$ 103, which corresponds to the molecular mass of $c$-C$_6$H$_5$CN.

\begin{figure*}
    \centering
    \includegraphics[width=11cm]{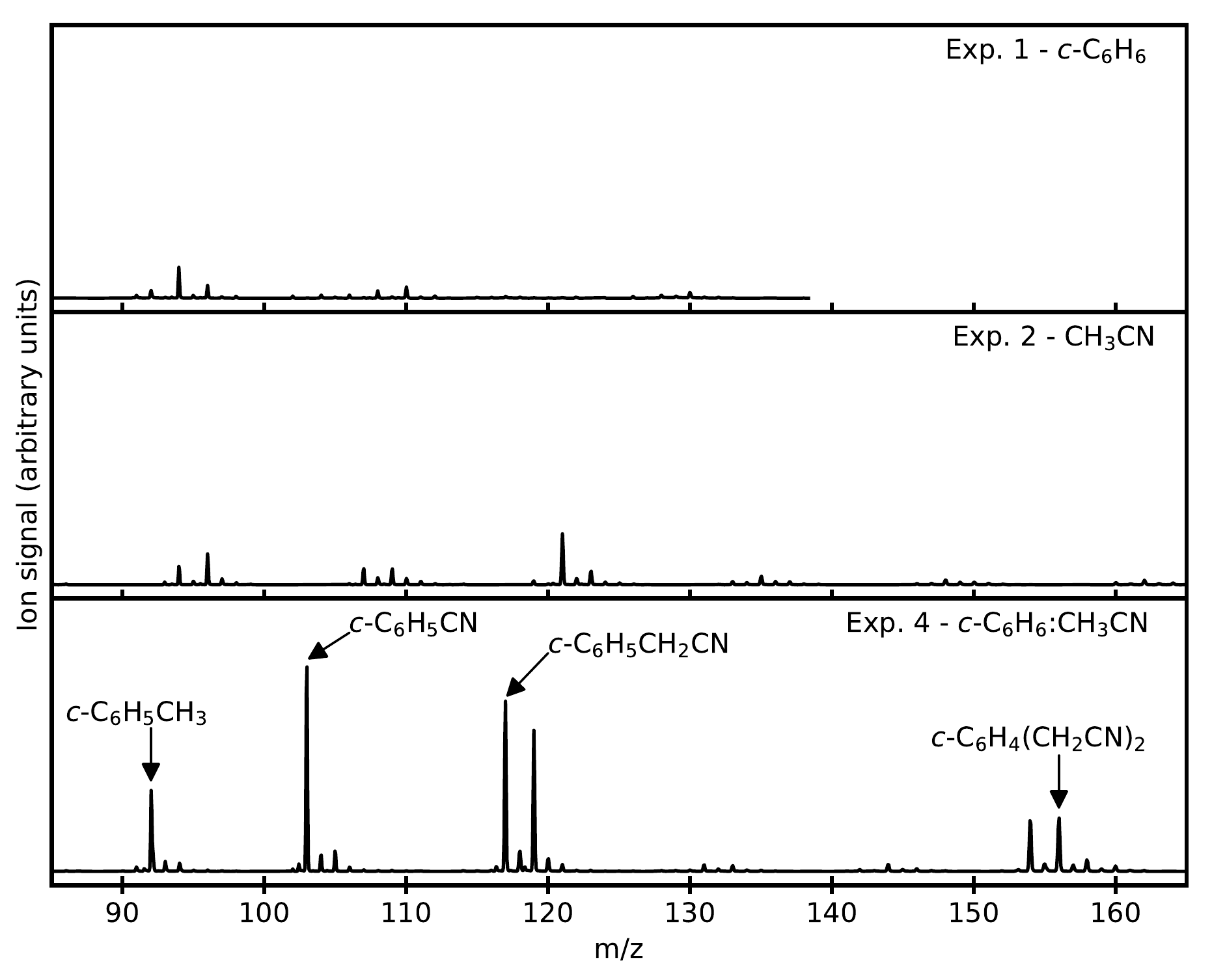}
    \caption{Mass spectrum (integrated ion signal during the TPD as a function of $m/z$) in the $m/z$ = 85$-$165 amu range of the desorbed ice samples after 2 keV electron irradiation for pure $c$-C$_6$H$_6$ and CH$_3$CN ices (Experiments 1 and 2, top and middle panels, respectively), and a $c$-C$_6$H$_6$:CH$_3$CN ice mixture (Exp. 4, bottom panel). 
    Peak assignments are indicated. 
    }
    \label{fig:ms}
\end{figure*}

Figure \ref{fig:ms} shows the mass spectrum (i.e., the ion signal integrated over the temperature axis as a function of $m/z$) of the desorbed ice samples in Experiments 1, 2, and 4 (top, middle, and bottom panels, respectively) in the $m/z$ = 85$-$165 amu range, where the peaks corresponding to benzene derivatives are expected to appear.  
%
The most intense peak 
detected in this range 
after irradiation of a $c$-C$_6$H$_6$:CH$_3$CN ice mixture is the $m/z$ $\approx$ 103 peak corresponding to $c$-C$_6$H$_5$CN (Fig. \ref{fig:ms}, bottom panel). 
This peak was not detected after 2 keV electron irradiation of ice samples containing only $c$-C$_6$H$_6$ or CH$_3$CN (top and middle panels of Fig. \ref{fig:ms}). 
Additional peaks were detected in Exp. 4 at $m/z$ $\approx$ 92, $m/z$ $\approx$ 117, and $m/z$  $\approx$ 156, corresponding to the molecular masses of toluene ($c$-C$_6$H$_5$CH$_3$), benzylnitrile ($c$-C$_6$H$_5$CH$_2$CN), and benzenediacetonitrile ($c$-C$_6$H$_4$(CH$_2$CN)$_2$). 
The peaks detected at $m/z$ $\approx$ 119 and $m/z$  $\approx$ 154 could not be securely identified without further experimentation,  
but the molecular masses are consistent with clusters formed by unreacted $c$-C$_6$H$_6$ and CH$_3$CN molecules, and with the loss of H$_2$ from a benzenediacetonitrile molecule, respectively. 
\begin{figure}[ht!]
    \centering
    \includegraphics[width=6.5cm]{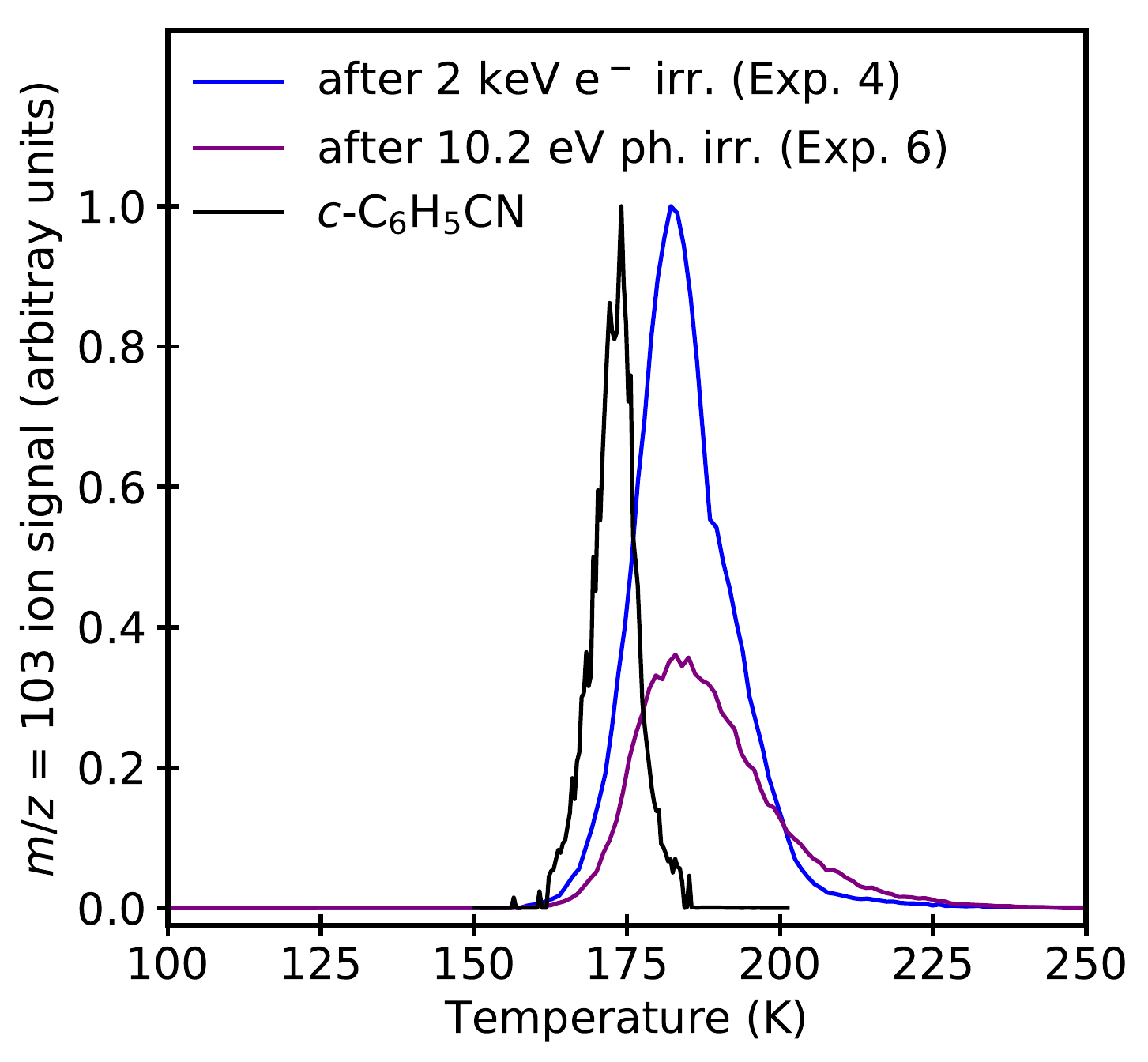}
    \caption{TPD curve for $m/z$ $\approx$ 103 of a pure $c$-C$_6$H$_5$CN ice sample (black, normalized to the value at the desorption peak) compared to that measured after irradiation of a 1:1 $c$-C$_6$H$_6$:CH$_3$CN ice mixture with 2 keV electrons (blue, Exp. 4 in Table \ref{tab:exp}) and Ly-$\alpha$ photons (purple, Exp. 6 in Table \ref{tab:exp}). The ion signal measured in Experiments 4 and 6 is normalized to the value at the desorption peak in Exp. 4.}
    \label{fig:tpd_binary}
\end{figure}

In order to confirm the assignment of the $m/z$ $\approx$ 103 peak to $c$-C$_6$H$_5$CN, Fig. \ref{fig:tpd_binary} shows the $m/z$ $\approx$ 103 TPD curve. 
The temperature of the desorption peak was 183 K, which 
is 9 K higher than the desorption peak temperature of a pure $c$-C$_6$H$_5$CN ice sample, also measured as part of this work and shown in Fig. \ref{fig:tpd_binary} as a reference. 
We note that subtle changes in the desorption temperatures are expected for molecules in ice mixtures compared to the values measured in pure ice samples. 
At the same time, we observed a decrease above 180 K in the absorbance of the two IR features assigned to $c$-C$_6$H$_5$CN in Fig. \ref{fig:ir} during the TPD of an irradiated $c$-C$_6$H$_6$:CH$_3$CN ice sample in an additional experiment (presented in Sect. \ref{sec:add} of the Appendix).
%


\subsection{m/z=103 isomers: benzonitrile and phenyl isocyanide formation}\label{sec:exp_binary_isomers}

While we consider the formation of $c$-C$_6$H$_5$CN upon 2 keV electron irradiation of a $c$-C$_6$H$_6$:CH$_3$CN ice mixture secure, based on the combination of consistent IR bands (Fig. \ref{fig:ir}) with the thermal desorption of a species with a molecular mass of 103 amu (Fig. \ref{fig:ms}) showing a consistent TPD curve (Fig. \ref{fig:tpd_binary}), we used the SPACE TIGER capabilities to explore whether there could be a substantial contribution to the $m/z$ $\approx$ 103 signal from any of the possible benzonitrile isomers.
%
%
%
Hydrogen loss and formation of CH$_3^.$ and CN$^.$ ice radicals were the most likely outcomes of CH$_3$CN ice irradiation with VUV photons in \citet{schwell08}. 
Likewise, hydrogen loss and fragmentation to C$_3$H$_3^.$ radicals were the dominant VUV irradiation products in $c$-C$_6$H$_6$ ice samples in \citet{kislov04}. 
From these assumptions, benzonitrile ($c$-C$_6$H$_5$CN) and phenyl isocyanide ($c$-C$_6$H$_5$NC) are the most likely products with a molecular mass of 103 amu that could be formed upon energetic processing of an ice sample containing $c$-C$_6$H$_6$ and CH$_3$CN.

\begin{figure}
    \centering
    \includegraphics[width=6.5cm]{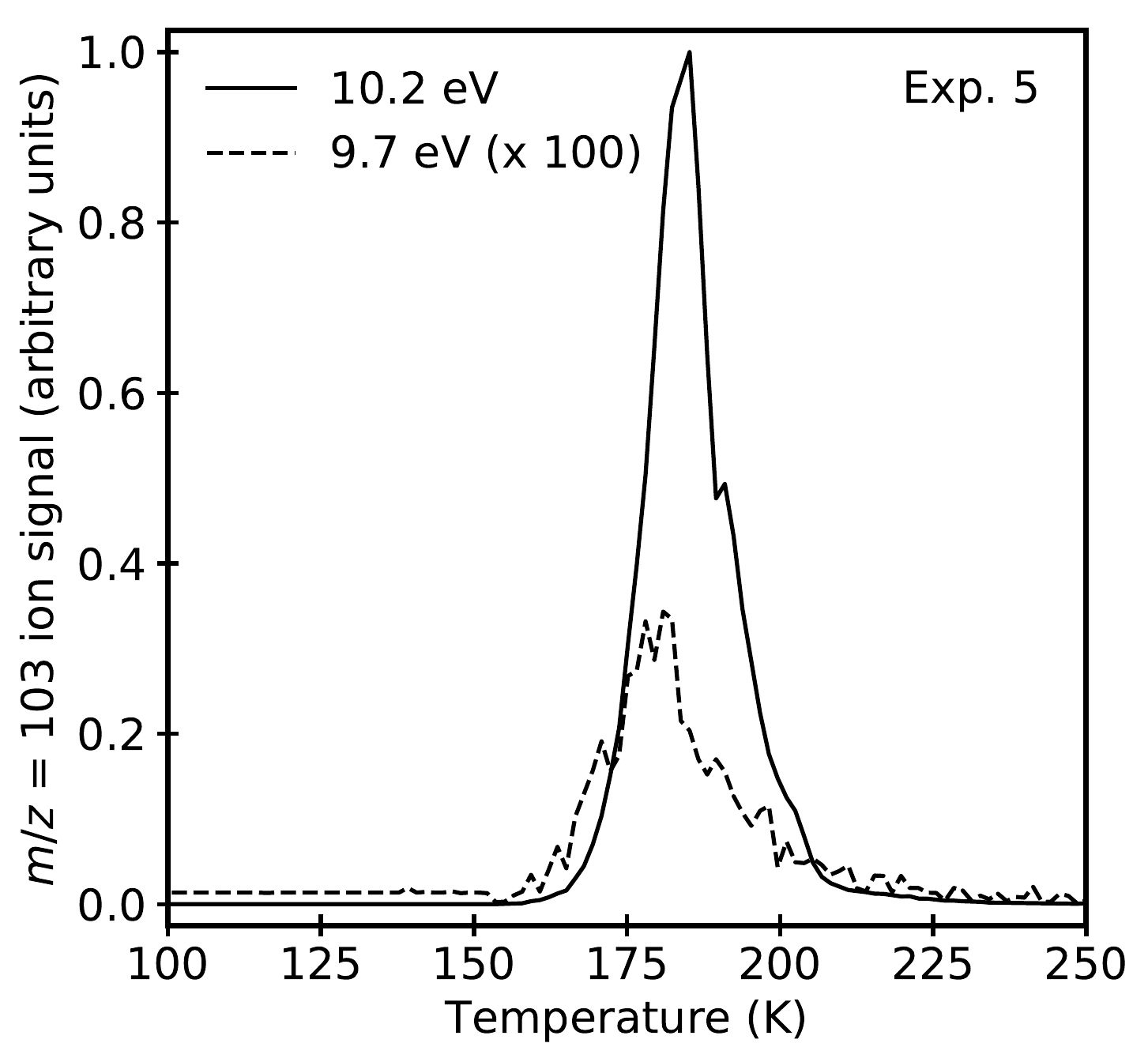}
    \caption{TPD curve for $m/z$ $\approx$ 103 of a 1:1 $c$-C$_6$H$_6$:CH$_3$CN ice mixture after irradiation with 2 keV electrons (Exp. 5 in Table \ref{tab:exp}) measured with an ionizing photon energy of 10.20 eV (solid line), that allowed detection of both benzonitrile and phenyl isocianide, and 9.70 eV (dashed line, multiplied by a factor of 100 for visualization purposes), that only allowed detection of phenyl isocyanide. The ion signal is normalized to the value at the desorption peak of the 10.2 eV TPD curve.
    }
    \label{fig:tpd_isomers}. 
\end{figure}

To constrain the contribution of phenyl isocyanide to the measured $m/z$ $\approx$ 103 signal, we used a lower ionizing photon energy of 9.70 eV in Exp. 5, in addition to the Ly-$\alpha$ ionizing radiation used in the rest of experiments. This energy falls in between the onset of the detected ion signals in the benzonitrile and phenyl isocyanide PIEs, which are 9.73 eV \citep[][although in our setup the measured ionization energy was 9.68 $\pm$ 0.01  eV, see Sect. \ref{sec:bn} of the Appendix]{araki96}, and 9.50 eV \citep[][]{young76}, respectively.
%
%
The two ionizing photon energies (10.20 eV and 9.70 eV) were alternatively used during the TPD of 
Exp. 5. 
Fig. \ref{fig:tpd_isomers} shows that the $m/z$ $\approx$ 103 ion signal measured with an ionizing photon energy of 10.20 eV was more than two orders of magnitude higher compared to that measured with the 9.70 eV ionizing energy. The latter was only able to ionize (and thus enable the detection of) phenyl isocyanide molecules, but not benzonitrile. 
This suggests that the dominant desorbing ice chemistry product in Experiments 3$-$5 with a molecular mass of 103 amu was benzonitrile.  

\subsection{Ly-$\alpha$ photon irradiaiton of a 1:1 $c$-C$_6$H$_6$:CH$_3$CN ice mixture}\label{sec:exp_binary_ph}

\begin{figure*}
    \centering
    \includegraphics[width=11cm]{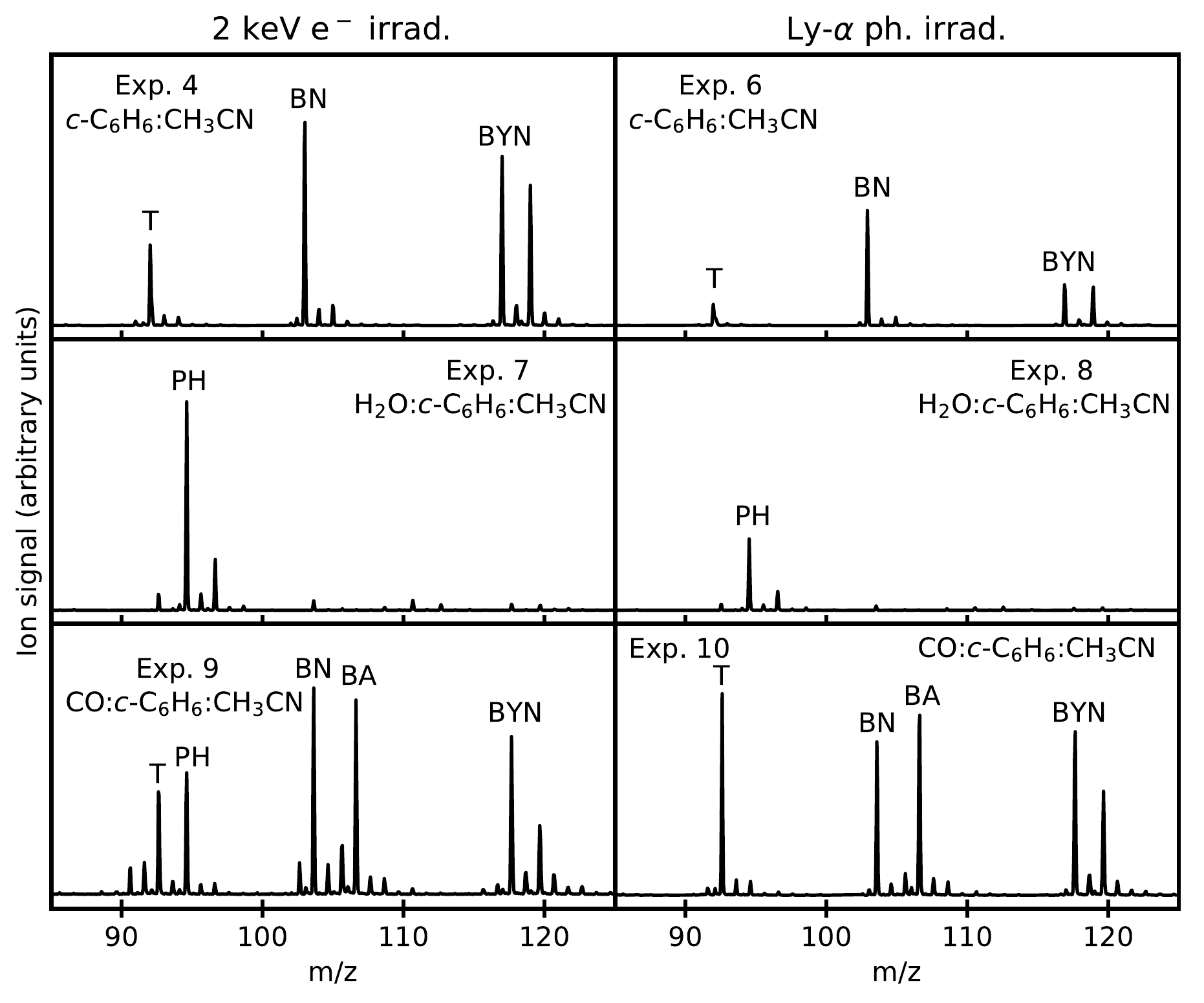}
    \caption{Mass spectrum (integrated ion signal during the TPD as a function of $m/z$) in the $m/z$ = 85$-$125 amu range of the desorbed ice samples after 2 keV electron (left panels) and Ly-$\alpha$ photon (right panels) irradiation of $c$-C$_6$H$_6$:CH$_3$CN, H$_2$O:$c$-C$_6$H$_6$:CH$_3$CN, and CO:$c$-C$_6$H$_6$:CH$_3$CN ice samples (top, middle and bottom panels, respectively). 
    Peak assignments are indicated with the corresponding initials: T (toluene, $c$-C$_6$H$_5$CH$_3$), PH (phenol, $c$-C$_6$H$_5$OH), BN (benzonitrile, $c$-C$_6$H$_5$CN), BA (benzaldehyde, $c$-C$_6$H$_5$CHO), and BYN (benzylnitrile, $c$-C$_6$H$_5$CH$_2$CN). 
    The ion signal scale is the same for panels in the same row. 
    }
    \label{fig:ms_matrix}
\end{figure*}

To evaluate the role of the ice energetic processing source, we ran one experiment where a 1:1 $c$-C$_6$H$_6$:CH$_3$CN ice sample was irradiated with Ly-$\alpha$ photons instead of 2 keV electrons (Exp. 6 in Table \ref{tab:exp}). 
The irradiated energy in Exp. 6 was of the same order as that of Experiments 3$-$5 (Table \ref{tab:exp}),  
and the results were similar to those obtained upon 2 keV electron irradiation of the same ice sample, as shown in the top left and right panels of Fig. \ref{fig:ms_matrix}. 
In particular, the corresponding $m/z$ $\approx$ 103 TPD curve (also shown in Fig. \ref{fig:tpd_binary}), 
presented an intensity on the same order of magnitude as the TPD curve measured after 2 keV electron irradiation of a similar ice sample in Exp. 4.  
This suggests that $c$-C$_6$H$_5$CN formation could reach similar yields upon both 2 keV electron and Ly-$\alpha$ photon irradiation of an ice sample containing $c$-C$_6$H$_6$ and CH$_3$CN molecules. 

\subsection{$c$-C$_6$H$_5$CN formation in H$_2$O and CO ice matrices}\label{sec:exp_ternary}

In order to further evaluate whether the formation of $c$-C$_6$H$_5$CN could take place under the conditions found in interstellar ice mantles, Experiments 7$-$10 in Table \ref{tab:exp} consisted in the irradiation with 2 keV electrons and Ly-$\alpha$ photons of 10:1:1 ice samples in which the $c$-C$_6$H$_6$ and CH$_3$CN molecules were embedded in a H$_2$O or CO ice matrix, simulating the chemistry in the interstellar H$_2$O-rich and CO-rich ice layers \citep{boogert15}. 
%
The purpose of these experiments was to qualitatively compare the relative production of $c$-C$_6$H$_5$CN and other benzene derivatives in different ice environments. 
The ice samples in Experiments 7$-$10 were exposed to similar electron and UV doses as in the simpler binary ice mixtures, and the formation of $c$-C$_6$H$_5$CN and other benzene derivatives was again explored by means of the Re-ToF MS during thermal desorption of the irradiated ice samples. 

The presence of O atoms in the H$_2$O-rich and CO-rich ice samples enabled additional reaction pathways that were not possible in a $c$-C$_6$H$_6$:CH$_3$CN binary mixture. 
This led to the formation and detection of O-bearing benzene derivatives, such as phenol ($c$-C$_6$H$_5$OH) and benzaldehyde ($c$-C$_6$H$_5$CHO). 
The left panels of Fig. \ref{fig:ms_matrix} show the mass spectra in the $m/z$ = 85$-$125 amu range of the desorbed ice samples after 2 keV electron irradiation of $c$-C$_6$H$_6$:CH$_3$CN (Exp. 4), H$_2$O:$c$-C$_6$H$_6$:CH$_3$CN (Exp. 7), and CO:$c$-C$_6$H$_6$:CH$_3$CN (Exp. 9) ice mixtures (top, middle, and bottom panels, respectively). 
The mass spectrum 
of the irradiated H$_2$O:$c$-C$_6$H$_6$:CH$_3$CN ice mixture (Fig. \ref{fig:ms_matrix}, middle left panel) was dominated by the $m/z$ $\approx$ 94 peak ($c$-C$_6$H$_5$OH), while the formation of $c$-C$_6$H$_5$CN ($m/z$ $\approx$ 103 peak) was negligible. 
On the other hand, the $m/z$ $\approx$ 103 peak was the most intense one in the mass spectrum of the 2 keV electron irradiated CO:$c$-C$_6$H$_6$:CH$_3$CN ice mixture (Fig. \ref{fig:ms_matrix}, bottom left panel), while the $m/z$ $\approx$ 92 ($c$-C$_6$H$_5$CH$_3$), $m/z$ $\approx$ 94, $m/z$ $\approx$ 106 ($c$-C$_6$H$_5$CHO), and $m/z$ $\approx$ 117 ($c$-C$_6$H$_5$CH$_2$CN) peaks corresponding to other benzene derivatives were about the same order of magnitude, as it was the case for the binary $c$-C$_6$H$_6$:CH$_3$CN ice mixture (Fig. \ref{fig:ms_matrix}, top left panel).

The right panels of Fig. \ref{fig:ms_matrix} show the mass spectra corresponding to the Ly-$\alpha$ photon irradiated ice samples (Experiments 6, 8, and 10). 
The results were similar to those obtained upon 2 keV electron irradiation of the same ice samples. 
In particular, the measured ion signals were on the same order of magnitude in electron and photon irradiated ices, as explained in Sect. \ref{sec:exp_binary_ph} for the $m/z$ $\approx$ 103 peak in irradiated $c$-C$_6$H$_6$:CH$_3$CN binary ice mixtures. 
In addition, the relative intensities of the peaks corresponding to different benzene derivatives did not show significant differences in the electron and photon irradiated ice samples with the same composition. 
The only exception was the negligible  $c$-C$_6$H$_5$OH formation in Ly-$\alpha$-irradiated CO-rich ices, whereas the intensity of the $m/z$ $\approx$ 94 peak in the 2 keV electron irradiated CO:$c$-C$_6$H$_6$:CH$_3$CN ice sample was on the same order as the other detected benzene derivatives. 
This is probably because Ly-$\alpha$ photons, unlike 2 keV electrons, were not able to effectively dissociate the CO molecules \citep{okabe78}, inhibiting the formation of $c$-C$_6$H$_5$OH molecules in the VUV photon irradiated ices.
In any case, these results suggest that the overall induced ice chemistry does not heavily depend on the irradiation source, as mentioned in Sect. \ref{sec:exp_binary_ph}.  

\section{Discussion}\label{sec:disc}

\subsection{Plausibility of $c$-C$_6$H$_5$CN formation in interstellar ice mantles}

%


Ice formation of $c$-C$_6$H$_5$CN in the laboratory was recently reported under conditions similar to those found in Titan's atmosphere \citep{mouzay21}. 
The experimental results presented in Sections \ref{sec:exp_binary_e} and \ref{sec:exp_binary_ph} show that formation of $c$-C$_6$H$_5$CN in the solid phase is also possible under the conditions found in the interior of the cold, dense clouds in the ISM, where ice mantles at very low temperatures containing $c$-C$_6$H$_6$ and a molecule with the cyano (C$\equiv$N) group (in this case, CH$_3$CN) could be exposed to secondary electrons \citep{bennett05} or UV photons \citep{shen04}. 

Formation of $c$-C$_6$H$_5$CN has also been observed when the reactants are diluted in ice analogs with astrophysically-relevant composition. 
In interstellar ices, $c$-C$_6$H$_6$ and CH$_3$CN (or any other cyanide) would most likely be 
embedded in a H$_2$O-rich or a CO-rich ice layer \citep{boogert15}. 
While in our experiments $c$-C$_6$H$_5$CN formation was quenched in a H$_2$O-rich ice sample, formation of this species was observed upon both 2 keV electron and a Ly-$\alpha$ photon irradiation of CO:$c$-C$_6$H$_6$:CH$_3$CN ice samples (Sect. \ref{sec:exp_ternary}). 
Therefore, our results indicate that solid-phase formation of $c$-C$_6$H$_5$CN from $c$-C$_6$H$_6$ and cyanide molecules in the ISM is a plausible pathway only if the reactants are primarily present on top of the ice mantles (which is not likely) or embedded in the CO-rich ice layer. 

\subsection{Relative contribution of the $c$-C$_6$H$_5$CN ice formation pathway}
One possible way to evaluate whether the proposed ice formation pathway could represent a significant contribution to the observed $c$-C$_6$H$_5$CN in the cold and dense regions of the ISM is to compare the product branching ratios of the different benzene derivatives found in irradiated CO:$c$-C$_6$H$_6$:CH$_3$CN ice samples with the observed abundance ratios in the ISM. 
In our experiments, five benzene derivatives were detected by the Re-ToF MS with comparable peak intensities (Fig. \ref{fig:ms_matrix}, bottom panels): toluene (C$_6$H$_5$CH$_3$), phenol ($c$-C$_6$H$_5$OH), benzonitrile ($c$-C$_6$H$_5$CN), benzaldehyde ($c$-C$_6$H$_5$CHO), and benzylnitrile ($c$-C$_6$H$_5$CH$_2$CN).  
%
Among these, phenol, benzonitrile, and benzaldehyde were searched for in the gas phase of the TMC-1 dense core \citep{mcguire18}.  
Since we have not calibrated the Re-ToF MS response for these molecules, we were not able to quantify the precise product branching ratios in our experiments, but we assumed (as a first approximation) that the Re-ToF MS response was comparable between different benzene derivatives. 
Based on the intensity of the corresponding peaks in the bottom panels of Fig. \ref{fig:ms_matrix}, we would expect similar gas-phase abundances for benzonitrile and benzaldehyde if these molecules were produced in the CO-rich layer of interstellar ices and subsequently desorbed to the gas-phase, while the abundance of phenol should be a factor of a few lower. 
Only the detection of benzonitrile was reported upon observations toward the TMC-1 dense core \citep[see Fig. 1 in][]{mcguire18}, with a column density of 4 $\times$ 10$^{11}$ cm$^{-2}$. 
Since no abundance upper limits were reported for phenol and benzaldehyde, we could not estimate lower limits for the abundance ratios. 
Therefore, the relative contribution of the $c$-C$_6$H$_5$CN ice formation pathway remains unconstrained based on the currently available observational data. 

%
%

\subsection{The $c$-C$_6$H$_5$CN/$c$-C$_6$H$_5$NC product branching ratio}
The results presented in Sect. \ref{sec:exp_binary_isomers} reveal that, among the two most likely C$_7$H$_5$N isomers that could be formed in irradiated ice samples containing $c$-C$_6$H$_6$ and CH$_3$CN, the formation of the -NC isomer phenyl isocyanide ($c$-C$_6$H$_5$NC) was negligible compared to that of benzonitrile ($c$-C$_6$H$_5$CN).
Previous theoretical and experimental studies have shown that the formation of phenyl isocyanide is also negligible in the gas phase, with less than 0.1\% relative abundance with respect to benzonitrile  \citep{balucani99,woon06,lee19}. 
This is probably due to a non-negligible energy barrier for the formation of the -NC isomer, compared to the barrierless formation of $c$-C$_6$H$_5$CN from $c$-C$_6$H$_6$ and CN. 
While benzonitrile has been observed toward several molecular clouds \citep{mcguire18,burkhardt21a}, no observational upper limits have been reported for the abundance of phenyl isocyanide in the ISM. In any case, the $c$-C$_6$H$_5$CN/$c$-C$_6$H$_5$NC relative abundance ratio would not be useful to evaluate the relative contribution of the solid- and gas-phase formation pathways, as both lead to a negligible formation of the -NC isomer.

\subsection{Electron- vs. photon-induced ice formation of $c$-C$_6$H$_5$CN}
%
In our experiments, irradiation 
with electrons or VUV photons led to $c$-C$_6$H$_5$CN ice formation yields on the same order of magnitude.  
In particular, the $m/z$ $\approx$ 103 ion signals measured during the TPD of electron and photon irradiated $c$-C$_6$H$_6$:CH$_3$CN ice samples were within a factor of 2$-$3 for comparable incident energies (Sect. \ref{sec:exp_binary_ph}). 
Similar results were found for CO:$c$-C$_6$H$_6$:CH$_3$CN ices (Sect. \ref{sec:exp_ternary}). 
We speculate that the small differences in the measured ion signals could be due to different absorbed energies in the 2 keV electron and Ly-$\alpha$ photon irradiated ice samples, leading to somewhat lower conversion yields per incident photon in the latter case. 
The amount of absorbed energy in electron and VUV photon irradiation experiments depends on the electron penetration depth and the VUV-photon absorption cross-sections, which have not yet been reported for $c$-C$_6$H$_6$ and CH$_3$CN ice. 
%

Since the energy dose experienced by the interstellar ice mantles in the interior of dense clouds from the cosmic-ray induced secondary UV field is thought to be similar in value to the energy contribution from the incoming cosmic rays \citep{moore01}, our results suggest that the contribution of both formation pathways (VUV-photon and cosmic-ray irradiation) to the ice formation of $c$-C$_6$H$_5$CN could be comparable in the ISM.
%

\section{Conclusions}\label{sec:conc}

We have presented the novel SPACE TIGER experimental setup designed to explore the physics and chemistry of interstellar ice mantles, and tested its capabilities by evaluating the plausibility of a $c$-C$_6$H$_5$CN ice formation pathway. 
SPACE TIGER includes two energy sources (a low energy electron source and a tunable VUV laser system) for the processing of ice samples. A Re-ToF MS coupled with the tunable laser system allows to discriminate between different isomers based on their ionization energies.  
Our experiments have revealed that:
\begin{enumerate}

    \item Formation of $c$-C$_6$H$_5$CN upon energetic processing of $c$-C$_6$H$_6$:CH$_3$CN ice samples is possible under the conditions found in cold, dense clouds. 
    
    \item $c$-C$_6$H$_5$CN formation was also observed in astrophysical ice analogs where $c$-C$_6$H$_6$ and CH$_3$CN were embedded in a CO ice matrix. However, the formation was quenched in a H$_2$O ice matrix where $c$-C$_6$H$_5$OH was the dominant ice chemistry product. 
     
    
    \item Electron and photon irradiation of ices containing $c$-C$_6$H$_6$ and CH$_3$CN led to comparable $c$-C$_6$H$_5$CN formation yields. Both pathways could thus equally contribute to the $c$-C$_6$H$_5$CN ice formation in the ISM. 
    
\end{enumerate} 
 
These results suggest that solid-phase formation of $c$-C$_6$H$_5$CN from $c$-C$_6$H$_6$ and cyanide molecules in the ISM is a plausible pathway only if the reactants are present on the ice surface or embedded in the CO-rich ice layer. 
Further evaluation of the relative contribution of the solid-phase $c$-C$_6$H$_5$CN formation pathway to the interstellar abundance of this species would require additional quantification of the product branching ratios, formation yields, and reaction kinetics, in combination with the use of theoretical models. 

\acknowledgments
This work was supported by a grant from the Simons Foundation (686302, K\"O) and an award from the Simons Foundation (321183FY19, K\"O).
The authors would like to acknowledge and appreciate the contribution of the late Pavlo Maksyutenko, not only to this work but also to the research group as a whole and to their lives. Our condolences go to his family and loved ones.

\appendix
\vspace{-8mm}
\section{Photoionization efficiency of gas-phase $c$-C$_6$H$_5$CN}\label{sec:bn}\label{sec:pie}

As explained in Sect. \ref{sec:tof}, the Re-ToF MS used in this work relies on single-photon ionization (SPI) of the thermally desorbed ice molecules prior to their detection. 
In order to select a working ionizing photon energy suitable for the detection of $c$-C$_6$H$_5$CN, 
we first measured the PIE of this species in the gas-phase by monitoring the $m/z$ = 103 ion signal detected with the Re-ToF MS as a function of the ionizing VUV photon energy. 

\begin{figure*}
    \centering
    \includegraphics[width=10cm]{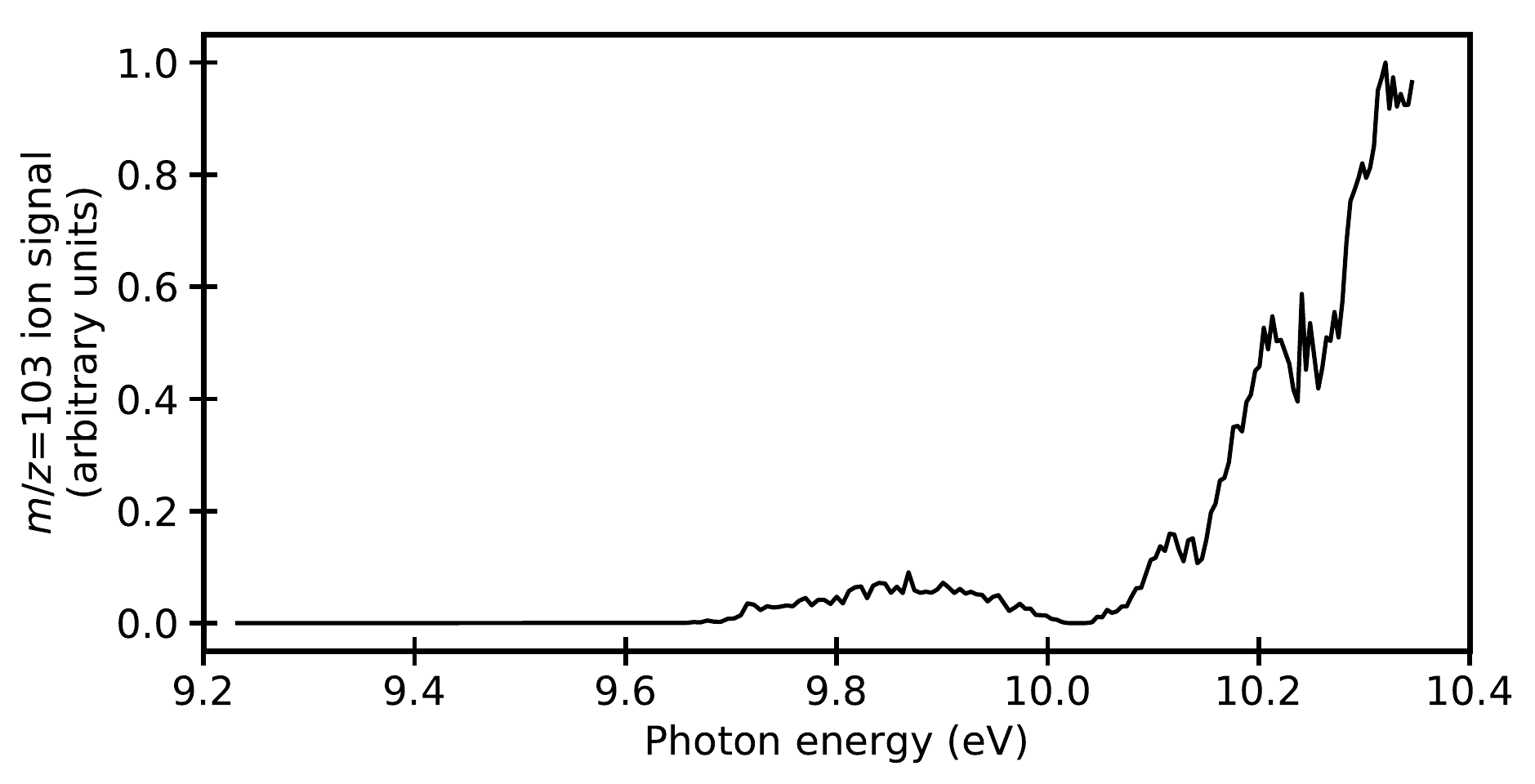}
    \includegraphics[width=10cm]{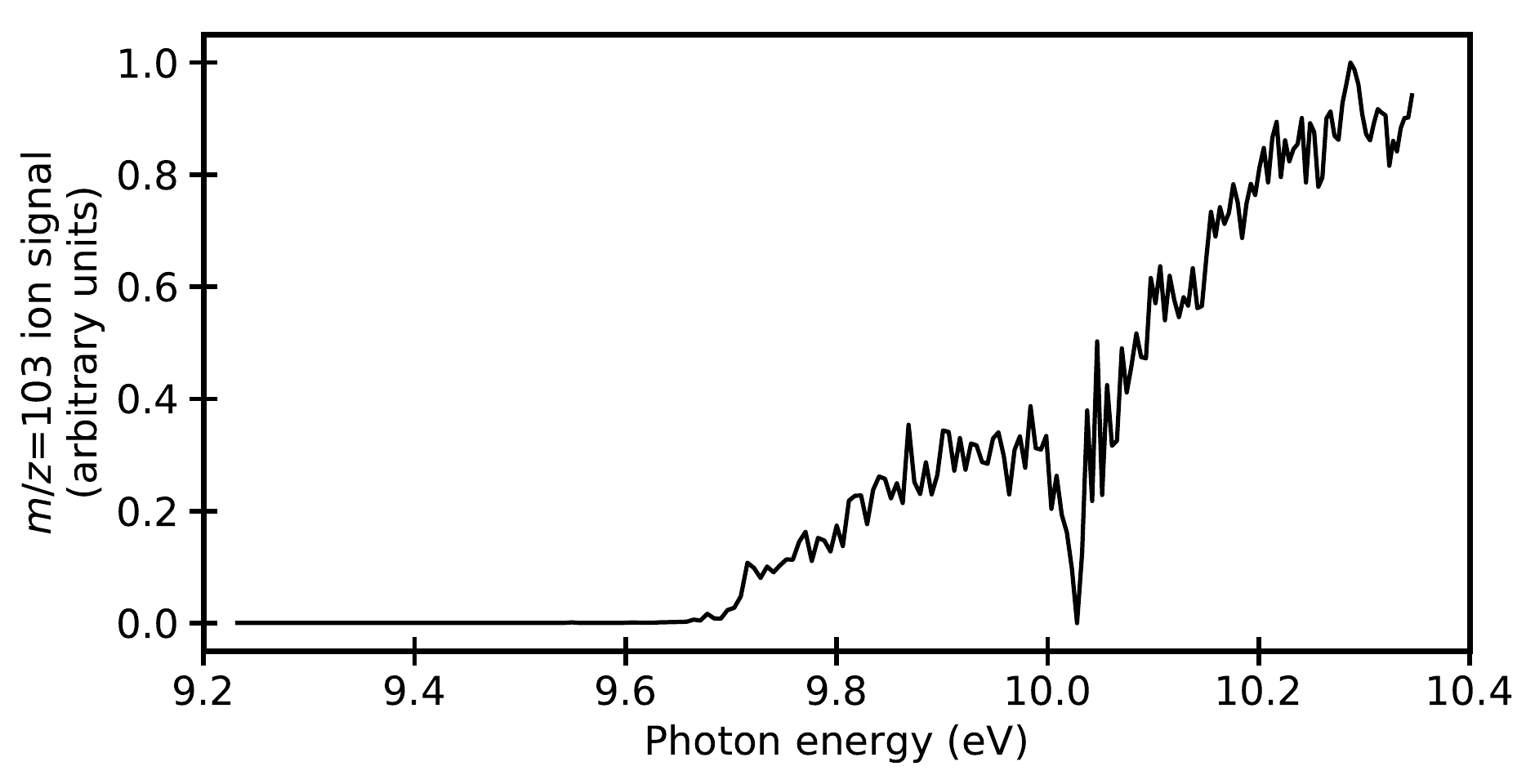}
    \caption{$c$-C$_6$H$_5$CN photoionization efficiency (PIE), measured as a) the $m/z$ $\approx$ 103 ion signal detected with the Re-ToF MS in SPACE TIGER as a function of the ionizing VUV photon energy when 1.8 $\times$ 10$^{-8}$ Torr of $c$-C$_6$H$_5$CN were present in the UHV chamber (\textit{top panel}, the ion signal is normalized to the maximum measured value); 
    b) the $m/z$ $\approx$ 103 ion signal in the same experiment, normalized to the VUV emission spectrum shown in Fig. \ref{fig:vuv} (\textit{bottom panel}).}
    \label{fig:pie}
\end{figure*}

We first introduced $c$-C$_6$H$_5$CN (liquid, 99.9\% purity, Sigma-Aldrich) in the gas mixing cylinder after applying three freeze-thaw-pump cycles with liquid N$_2$. 
%
%
$c$-C$_6$H$_5$CN was subsequently admitted into the UHV chamber up to a pressure of 1.8$\times$10$^{-8}$ Torr (while keeping the copper substrate at room temperature). 
The OPO was then set to linearly scan from 410 nm to 650 nm in 200 steps (90 laser pulses per step), producing VUV photons with an energy between 9.23 eV and 10.35 eV. 
The top panel of Fig. \ref{fig:pie} shows the $m/z$ $\approx$ 103 ion signal corresponding to the detection of $c$-C$_6$H$_5$CN molecules, measured by the Re-ToF MS as a function of the ionizing VUV photon energy.  
The ion signal measured by the Re-ToF MS is affected by the VUV photon flux of the ionizing radiation, which varies with the wavelength (Fig. \ref{fig:vuv}). 
Therefore, the bottom panel of Fig. \ref{fig:pie} shows the $m/z$ = 103 ion signal normalized to the VUV emission spectrum shown in Fig. \ref{fig:vuv}. This should be a more representative measurement of the $c$-C$_6$H$_5$CN PIE, as it is independent from the experimental setup. 
%
%
%
The onset of the detected ion signal at 9.68 $\pm$ 0.01 eV corresponds to the 
minimum ionizing photon energy that allows the detection of $c$-C$_6$H$_5$CN by the Re-ToF MS. This energy is close to the literature value for the gas-phase $c$-C$_6$H$_5$CN ionization energy \citep[9.73 $\pm$ 0.01  eV,][]{araki96}. 
%
We note that the drop in the $m/z$ = 103 ion signal observed around 10.03 eV in Fig. \ref{fig:pie} corresponds to the drop in the VUV photon flux at the same wavelength observed in Fig. \ref{fig:vuv}. 
%
In the subsequent experiments, we used a photon energy for Re-ToF MS detection of 10.20 eV (corresponding to the Ly-$\alpha$ wavelength) well above the PIE threshold.

\section{IR spectroscopy during TPD of a 2 keV electron irradiated $c$-C$_6$H$_6$:CH$_3$CN ice mixture}\label{sec:add}

\begin{figure*}
    \centering
    \includegraphics[width=14cm]{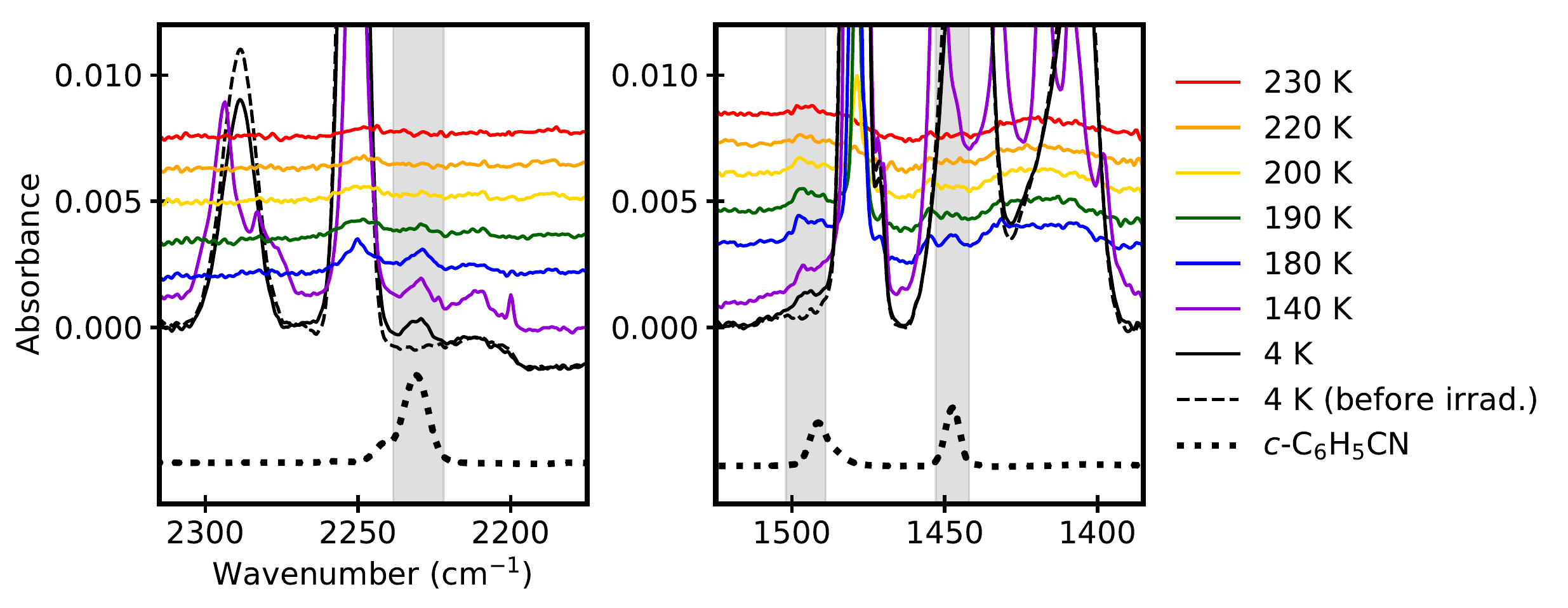}
    \caption{IR spectra collected in reflection-absorption mode during the TPD after irradiation with 2 keV electrons of a 1:1 $c$-C$_6$H$_6$:CH$_3$CN ice sample. 
    IR features assigned to $c$-C$_6$H$_5$CN at 2230 cm$^{-1}$, 1495 cm$^{-1}$, and 1447 cm$^{-1}$ are highlighted in the left and right panels.
   The IR spectrum of a pure $c$-C$_6$H$_5$CN ice (black dashed line) is also shown as a reference.
   }
    \label{fig:add}
\end{figure*}

In order to confirm the assignment of the 2230 cm$^{-1}$ and 1495 cm$^{-1}$ IR features observed in the bottom panel of Fig. \ref{fig:ir} to $c$-C$_6$H$_5$CN, we performed an additional 2 keV electron irradiation experiment with a 1:1 $c$-C$_6$H$_6$:CH$_3$CN ice sample. 
The same procedure as for Experiments 3$-$5 in Table \ref{tab:exp} was followed with the exception that, during the TPD of the irradiated ice mixture, we collected IR spectra of the solid sample in reflection-absorption mode instead of detecting the desorbing molecules with the Re-ToF MS. 

Fig. \ref{fig:add} shows the evolution of the 2230 cm$^{-1}$ (left panel) and 1495 cm$^{-1}$ (right panel) features between 140 K and 230 K, along with other CH$_3$CN (2290 cm$^{-1}$, 2250 cm$^{-1}$, 2200 cm$^{-1}$, 1450 cm$^{-1}$, 1405 cm$^{-1}$) and C$_6$H$_6$ (2210 cm$^{-1}$, 1480 cm$^{-1}$) bands observed in the IR spectrum of the irradiated ice mixture. 
A change in the structure of the irradiated ice sample between 4 K and 140 K was evidenced by a change in the profile of the CH$_3$CN (and, to a lower extent, C$_6$H$_6$) features.   
The majority of the (unreacted) C$_6$H$_6$ and CH$_3$CN molecules desorbed between 140 K and 180 K, leading to a significant decrease in the absorbance of the corresponding IR bands. 
This enabled the detection of a small feature at 1447 cm$^{-1}$ in the right panel of Fig. \ref{fig:add}. 
This feature was overlapped with the 1450 cm$^{-1}$ CH$_3$CN IR band, and could also correspond to $c$-C$_6$H$_5$CN according to the pure $c$-C$_6$H$_5$CN ice spectrum also shown in Fig. \ref{fig:add} as a reference. 
Between 180 K and 200 K a significant decrease in the absorbance of the 2230 cm$^{-1}$, 1495 cm$^{-1}$, and 1447 cm$^{-1}$ IR features was observed. 
This temperature range is consistent with the desorption 
observed between $\sim$160 K and $\sim$210 K 
in the $m/z$ $\approx$ 103 TPD curve (Fig. \ref{fig:tpd_binary}, also assigned to $c$-C$_6$H$_5$CN), and confirms the assignment of the 2230 cm$^{-1}$, 1495 cm$^{-1}$, and 1447 cm$^{-1}$ IR features to $c$-C$_6$H$_5$CN.
We note that the 1495 cm$^{-1}$ feature showed a broad profile, with the higher-frequency wing completely disappearing between 200 K and 220 K (along with the remaining of the 1480 cm$^{-1}$ C$_6$H$_6$ band).  
While this is also consistent with the $c$-C$_6$H$_5$CN desorption temperature mentioned above, we cannot rule out a contribution from a different species to this particular feature.


\begin{thebibliography}{}
\bibitem[Abplanalp et al.(2019)]{abplanalp19} Abplanalp, M.J., Frigge, R., \& Kaiser, R.I.\ 2019, SciAdv, 5, 10, 5841
\bibitem[Allamandola et al.(1985)]{allamandola85} Allamandola, L.J., Tielens, A.G.G.M., \& Barker, J.R.\ 1985, \apj, 290, L25
\bibitem[Araki et al.(1996)]{araki96}Araki, M., Sato, S.-i., \& Kimura, K.\ 1996, The Journal of Physical Chemistry, 100 (25), 10542
\bibitem[Balucani et al.(1999)]{balucani99} Balucani, N., Asvany, O., Chuang, A.H.H., et al.\ 1999, JChPh, 111, 7457
\bibitem[Bennett et al.(2005)]{bennett05} Bennett, C.~J., Jamieson, C.~S., Osamura, Y., \& Kaiser, R.~I.\ 2005, \apj, 624, 1097
\bibitem[Bern\'e et al.(2015)]{berne15} Bern\'e, O., Montillaud, J., \& Joblin, C.\ 2015, A\&A, 577, A133
\bibitem[Boesl et al.(1991)]{boesl91} Boesl, U., Weinkauf, R., \& Schlag, E.W.\ 1991, IJMS, 112, 121
\bibitem[Boogert et al.(2015)]{boogert15} Boogert, A.~C.~A., Gerakines, P.~A., \& Whittet D.~C.~B.\ 2015, ARA\&A, 53, 541
\bibitem[Burkhardt et al.(2021a)]{burkhardt21a} Burkhardt, A.M., Loomis, R.A., Shingledecker, C.N., et al.\ 2021a, Nature Astronomy, 5, 181
\bibitem[Burkhardt et al.(2021b)]{burkhardt21} Burkhardt, A.M., et al.\ 2021b, \apj, submitted
\bibitem[Cecchi-Pestellini \&Aiello(1992)]{cecchi92} Cecchi-Pestellini, C., \& Aiello, S.\ 1992, \mnras, 258, 125
\bibitem[Cernicharo(2004)]{cernicharo04}Cernicharo, J.\ 2004, \apj, 608, L41
\bibitem[Cernicharo et al.(2021)]{cernicharo21}Cernicharo, J., Ag\'undez, M., Cabezas, C., et al.\ 2021, \apj, 649, L15
\bibitem[Cherchneff et al.(1992)]{cherchneff92} Cherchneff, I., Barker, J.R., \& Tielens, A.G.G.M.\ 1992, \apj, 401, 269
\bibitem[Chiar et al.(2013)]{chiar13}Chiar, J.E., Tielens, A.G.G.M., Adamson, A.J., \& Ricca, A.\ 2013, \apj, 770, 78
\bibitem[Cooke et al.(2020)]{ilsa20}Cooke, I.R., Gupta, D., Messinger, J.P., \& Sims, I.R.\ 2020, \apjl, 891, L41
\bibitem[Dwek et al.(1997)]{dwek97}Dwek, E., Arendt, R.G., Fixsen, D.J., et al.\ 1997, ESC, 3, 1109
\bibitem[Hamon et al.(2000)]{hamon00}Hamon, S., Le Picard, S.D., Canosa, A., Rowe, B.R., \& Smith, I.W.M.\ 2000, JChPh, 109, 3882 
\bibitem[Hanna et al.(2009)]{hanna09} Hanna, S.J., Campuzano-Jost, P., Simpson, E.A., et al.\ 2009, IJMS, 279, 2$-$3, 134
\bibitem[Joblin \& Cernicharo(2018)]{joblin18}Joblin, C., \& Cernicharo, J.\ 2018, Science, 359, 156
\bibitem[Jones et al.(2011)]{jones11} Jones, B.~M., Bennett, C.~J., \& Kaiser, R.\ 2011, \apj, 734, 78
\bibitem[Kaiser et al.(2015)]{kaiser15}Kaiser, R.I., Parker, D.S.N., \& Mebel, A.M., 2015, ARPC, 66, 43
\bibitem[Kislov et al.(2004)]{kislov04}Kislov, V., Nguyen, T., Mebel, A., Lin, S., \& Smith, S.\ 2004, The Journal of chemical physics, 120 (15), 7008.
\bibitem[Kolesnikov\'a et al.(2013)]{kolesnikova13} Kolesnikov\'a, L., Daly, A.M., Alonso, J.L., Tercero, B., \& Cernicharo, J.\ 2013, J. Mol. Spectrosc., 289, 13
\bibitem[Lee et al.(2021)]{lee21} Lee, K.L.K., Changala, P.B., Loomis, R.A., et al.\ 2021, \apjl, 910, L2
\bibitem[Lee et al.(2019)]{lee19} Lee, K.L.K., McGuire, B.A., \& McCarthy, M.C.\ 2019, PCCP, 21, 2946
\bibitem[L\'eger \& Puget(1984)]{leger84}L\'eger, A., \& Puget, J.L.\ 1984, A\&A, 137, L5
\bibitem[Low at al.(1984)]{low84}Low, F.J., Beintema, D.A., Gautier, T.N., et al.\ 1984, \apj, 278, L19
\bibitem[Marangos et al.(1990)]{marangos90} Marangos, J.P., Shen, N., Ma, H., Hutchinson, M.H.R., \& Connerade, J.P.\ 1990, JOSA B, 7, 7, 1254
\bibitem[Mart\'in-Dom\'enech et al.(2020)]{martin20} Mart\'in-Dom\'enech, R., Maksiutenko, P., \"Oberg, K.I., \& Rajappan, M.\ 2020, \apj, 902, 2, 116 
\bibitem[Mart\'inez et al.(2020)]{martinez20} Mart\'inez, L., Santoro, G., Merino, P., et al.\ 2020, Nat. Astron., 4, 97
\bibitem[McCarthy et al.(2020)]{mccarthy20}McCarthy, M.C., Lee, K.L.L., Loomis, R.A., et al.\ 2020, Nature Astronomy, 1
\bibitem[McGuire et al.(2018)]{mcguire18} McGuire, B., Burkhardt, A.M., Kalenskii, S. et al.\ 2018, Science, 359, 202
\bibitem[McGuire et al.(2021)]{mcguire21} McGuire, B., Loomis, R.A., Burkhardt, A.M., et al.\ 2021. Science, 371, 6535, 1265
\bibitem[Moore et al.(2001)]{moore01} Moore, M.H., Hudson, R.L., \& Gerakines, P.A.\ 2001, Spectrochimica Acta Part A, 57, 843
\bibitem[Mouzay et al.(2021)]{mouzay21} Mouzay, J, Henry, K., Couturier-Tamburelli. I., et al.\ 2021, Icarus, 368, 114595
\bibitem[\"Oberg et al.(2009)]{oberg09}\"Oberg, K.I., van Dishoeck, E.F., \& Linnartz, H.\ 2009, A\&A, 496, 281
\bibitem[Okabe(1978)]{okabe78} Okabe, H.\ 1978, Photochemistry of small molecules (New York: John Wiley \& Sons)
\bibitem[Pilleri et al.(2015)]{pilleri15}Pilleri, P., Joblin, C., Boulanger, F., \& Onaka, T.\ 2015, A\&A, 577, A16
\bibitem[Schneiderman et al.(2022)]{tajana22} Schneiderman, T., et al. \ 2022, \textit{in prep.}
\bibitem[Schwell et al.(2008)]{schwell08}Schwell, M., Jochims, H.-W., Baumg\"artel, H., \& Leach, S.\ 2008, Chemical Physics, 344 (1), 164.
\bibitem[Shen et al.(2004)]{shen04} Shen, C.~J., Greenberg, J.~M., Schutte, W.~A., \& van Dishoeck, E.~F.,\ 2004, A\&A, 415, 203
\bibitem[Tielens \& Charnley(1997)]{tielens97} Tielens, A.G.G.M., \& Charnley, S.B.\ 1997, in Planetary and Interstellar Processes Relevant to the Origins of Life, ed. D.C.B. Whittet (Berlin:Springer), 23
\bibitem[Trevitt et al.(2010)]{trevitt10} Trevitt, A.J., Goulay, F., Taatjes, C.A., Osborn, D.L., \& Leone, S.R.\ 2010, JPCA, 114, 1749
\bibitem[Woods et al.(2002)]{woods02} Woods, P.M., Millar, T.J., \& Zijlstra, A.A.\ 2002, \apj, 574, L167
\bibitem[Woon(2006)]{woon06}Woon, D.E.\ 2006, CP, 331, 67
\bibitem[Young \& Cheng(1976)]{young76} Young, V. Y. \& Cheng, K. L.\ 1976, Journal of Electron Spectroscopy and Related Phenomena, 9 (3), 317.

\end{thebibliography}
\end{document}